\documentclass[a4paper, accepted=2021-03-09]{quantumarticle}

\pdfoutput=1

\usepackage{authblk}
\usepackage{amsmath, amssymb}

\usepackage[utf8]{inputenc}
\usepackage[T1]{fontenc}

\usepackage{tikz}
\usetikzlibrary{positioning}
\usetikzlibrary{decorations.pathreplacing}

\usepackage{booktabs} 
\usepackage{tabularx}
\usetikzlibrary{calc}
\usepackage{pgfplots}

\usepackage[breaklinks=true,colorlinks=true,linkcolor=blue,urlcolor=blue,citecolor=blue]{hyperref}

\usepackage{titlecaps}
\usepackage{mathtools}

\usepackage[citestyle=numeric-comp, sorting=none, maxnames=4, backend=bibtex]{biblatex}

\pgfplotsset{compat=newest, width=0.72\textwidth, scaled y ticks = false, height=0.6*\axisdefaultheight, scale only axis}

\usepackage{graphicx}
\usepackage{subcaption}
\usepackage{cleveref}

\usepackage{xcolor}

\usepackage{algorithmicx}
\usepackage{algorithm}
\usepackage{algpseudocode}
\usepackage{titlecaps}

\usepgfplotslibrary{groupplots}
\usepgfplotslibrary{dateplot}

\renewbibmacro*{doi+eprint+url}{%
    \printfield{doi}%
    \newunit\newblock%
    \iffieldundef{doi}{%
        \usebibmacro{eprint}}%
        {}%
    }

\DeclarePairedDelimiter{\floor}{\lfloor}{\rfloor} 
\DeclarePairedDelimiter{\ket}{|}{\rangle}

\newcommand{\CX}{\textsf{CX}}
\newcommand{\CZ}{\textsf{CZ}}
\renewcommand{\H}{\textsf{H}}
\newcommand{\I}{\textsf{I}}
\newcommand{\Y}{\textsf{Y}}
\newcommand{\Z}{\textsf{Z}}
\newcommand{\X}{\textsf{X}}
\newcommand{\RZ}{\textsf{RZ}}
\newcommand{\RX}{\textsf{RX}}
\newcommand{\U}{{\sf U}}
\newcommand{\Uone}{{\sf U_1}}
\newcommand{\Utwo}{{\sf U_2}}
\newcommand{\Uthree}{{\sf U_3}}
\newcommand{\SWAP}{\textsf{SWAP}}

\newcommand{\tket}{\textsf{t}$\ket{\mathsf{ket}}$}
\newcommand{\pytket}{\textsf{pytket}}
\newcommand{\qiskit}{\textsf{Qiskit}}
\newcommand{\IBMnoise}{\textsf{noise-aware Qiskit}}
\newcommand{\IBM}{\textsf{noise-unaware Qiskit}}
\newcommand{\CQCnoise}{\textsf{noise-aware pytket}}
\newcommand{\CQC}{\textsf{noise-unaware pytket}}
\newcommand{\NONE}{\textsf{only pytket routing}}
\newcommand{\noisefree}{Noise-Free}
\newcommand{\figcap}[1]{}

\newcommand{\shallow}{shallow circuits}
\newcommand{\Shallow}{Shallow Circuits}
\renewcommand{\square}{square circuits}
\newcommand{\Square}{Square Circuits}
\newcommand{\deep}{deep circuits}
\newcommand{\Deep}{Deep Circuits}
\newcommand{\stack}{quantum computing stack}
\newcommand{\supremacy}{quantum computational supremacy}
\newcommand{\defpar}{\paragraph{Definitions and Related Results}}
\newcommand{\motpar}{\paragraph{Discussion}}
\newcommand{\quantimp}{\paragraph{Evaluating Stack Performance}}
\newcommand{\classver}{\paragraph{Empirical Estimation From Samples}}
\newcommand{\poly}[1]{\mathrm{poly} \brac{#1}}

\newcommand{\HOG}{Heavy Output Generation Benchmarking}
\newcommand{\XE}{Cross-Entropy Benchmarking}

\newcommand{\simins}{Insights from Classical Simulation}
\newcommand{\fullstack}{Full Stack Benchmarking}
\newcommand{\appmotiv}{Application Motivated Benchmarks}

\newcommand{\inout}[3]{
	\vspace{7pt}
	\hspace*{\algorithmicindent} \textbf{Input:} #1 \\
	\hspace*{\algorithmicindent} \textbf{Worst case depth:} #2 \\
	\hspace*{\algorithmicindent} \textbf{Output:} #3 

	\hrulefill
	}

\algdef{SE}[DOWHILE]{Do}{doWhile}{\algorithmicdo}[1]{\algorithmicwhile\ #1}

\newcolumntype{b}{X}
\newcolumntype{m}{>{\hsize=.75\hsize}X}
\newcolumntype{s}{>{\hsize=.5\hsize}X}

\newtheorem{problem}{Problem}

\newtheorem{theorem}{Theorem}

\newtheorem{definition}{Definition}

\newcommand{\bra}[1]{\left\langle #1 \right|}
\renewcommand{\mod}[1]{\left| #1 \right|}
\newcommand{\brac}[1]{\left( #1 \right)}
\newcommand{\sqrbrac}[1]{\left[ #1 \right]}
\newcommand{\curlbrac}[1]{\left\{ #1 \right\}}

\newcommand{\lone}{$\ell_1$-norm distance}

\newcommand{\melb}{\textsf{ibmq\_16\_melbourne}}
\newcommand{\york}{\textsf{ibmqx2}}
\newcommand{\our}{\textsf{ibmq\_ourense}}
\newcommand{\sing}{\textsf{ibmq\_singapore}}

\newcommand{\secref}[1]{Section \ref{#1}}

\newcommand{\thmref}[1]{Theorem \ref{#1}}
\newcommand{\equref}[1]{equation (\ref{#1})}
\newcommand{\tabref}[1]{Table \ref{#1}}

\renewcommand{\algref}[1]{Algorithm \ref{#1}}
\newcommand{\figref}[1]{Figure \ref{#1}}
\newcommand{\appref}[1]{Appendix \ref{#1}}
\newcommand{\probref}[1]{Problem \ref{#1}}

\newcommand{\hog}[2]{\mathrm{HOG} \brac{#1 , #2}}

\newcommand{\dist}[1]{\mathcal{#1}}

\newcommand{\xedef}{
  The \emph{cross-entropy} between two probability distributions $\dist{D}$
  and $\dist{D'}$is
  \begin{equation}
    \mathrm{CE} \brac{\dist{D}, \dist{D'}} = \sum_{x \in \curlbrac{0
    , 1}^{n}} \dist{D} \brac{x} \log \brac{\frac{1}{\dist{D'}
    \brac{x}}}.
  \end{equation}}

\newcommand{\crule}[1]{[\textcolor{#1}{\rule{0.2cm}{0.2cm}}]}

\definecolor{2}{HTML}{ffbe0b}
\definecolor{3}{HTML}{3b83cc}
\definecolor{4}{HTML}{97db4f}
\definecolor{5}{HTML}{ff006e}
\definecolor{6}{HTML}{ec7505}
\definecolor{7}{HTML}{808c9c}

\definecolor{ibmq_16_melbourne}{HTML}{3D77A4}
\definecolor{ibmqx2}{HTML}{77BB9A}
\definecolor{ibmq_ourense}{HTML}{EFD8D2}
\definecolor{ibmq_singapore}{HTML}{F26419}
\definecolor{ideal}{HTML}{AFB7C0}
\definecolor{CQC}{HTML}{FF595E}
\definecolor{CQC-noise}{HTML}{FFCA3A}
\definecolor{CQC-noise_v4}{HTML}{ffffff}
\definecolor{IBM}{HTML}{38A3E5}
\definecolor{IBM-noise}{HTML}{896BB3}
\definecolor{NONE}{HTML}{8AC926}
\definecolor{all}{HTML}{ffffff}

\graphicspath{ {./imag/} }

\title{Application-Motivated, Holistic Benchmarking of a Full Quantum Computing Stack}

\author[1,2]{Daniel Mills}
\email{daniel.mills@cambridgequantum.com}
\orcid{0000-0001-5902-3774}
\author[2]{Seyon Sivarajah}
\author[3]{Travis L. Scholten}
\orcid{0000-0001-9334-4248}
\author[2,4]{Ross Duncan}

\affil[1]{University of Edinburgh, 10 Crichton Street, Edinburgh EH8 9AB, UK}
\affil[2]{Cambridge Quantum Computing Ltd, 9a Bridge Street, Cambridge, CB2 1UB, UK}
\affil[3]{IBM T.J. Watson Research Center, Yorktown Heights, NY 10598, USA}
\affil[4]{University of Strathclyde, 26 Richmond Street, Glasgow, G1 1XH, UK}

\begin{document}

\maketitle

\begin{abstract}
	Quantum computing systems need to be benchmarked in terms of practical
	tasks they would be expected to do. Here, we propose 3 ``application-
	motivated'' circuit classes for benchmarking: deep (relevant for state
	preparation in the variational quantum eigensolver algorithm), shallow
	(inspired by IQP-type circuits that might be useful for near-term
	quantum machine learning), and square (inspired by the quantum volume
	benchmark).  We quantify the performance of a quantum computing system
	in running circuits from these classes using several figures of merit,
	all of which require exponential classical computing resources and a
	polynomial number of classical samples (bitstrings) from the system. We
	study how performance varies with the compilation strategy used and the
	device on which the circuit is run. Using systems made available by IBM
	Quantum, we examine their performance, showing that noise-aware
	compilation strategies may be beneficial, and that device connectivity
	and noise levels play a crucial role in the performance of the system
	according to our benchmarks.
\end{abstract}

\newpage

\tableofcontents

\section{Introduction} 
\label{sec:introduction}

The evolution of quantum computers from experimental devices comprising a
handful of qubits, towards general-purpose, programmable, commercial-grade
systems \cite{qiskit, Karalekas_2020, Arute2019, devitt2016performing,
Debnath2016} necessitates new techniques for characterizing them. Quantum
characterization, validation, and verification (QCVV) protocols to detect,
diagnose, and quantify errors in quantum computers originally focused on
properties of one or several qubits (e.g., $T_{1}$ and $T_{2}$ times, gate
error rates, state preparation fidelity, etc). Such quantities measure the form
and magnitude of noise present in the system, but are only \emph{proxy}
measures of device performance when implementing computations.

This can be addressed through ``holistic'' benchmarks, which stress test a
quantum computing system in its \emph{entirety}, and not just individual
components. ``Holistic benchmarking'' of a quantum computing system has
commonly referred to benchmarking the physical implementation of a collection
of qubits, without referring to the computational task they would perform. This
is most useful when testing \emph{physical properties} of a collection of
qubits.\footnote{A simple example is crosstalk detection, where the output of
the benchmarking could be a table of coupling values between all connected
qubits.}

A complementary view (taken in this work) is that holistic benchmarks test the
\emph{computational capabilities} of the complete system. Under this view, the
entire compute \emph{stack} -- qubits, compilation strategy, classical control
hardware, etc. -- should be benchmarked collectively
\cite{murali2019comparicon}. This captures the performance of the system as an
integrated unit, giving a better indication of the system's performance in
practice.

Holistic benchmarks, while useful for comparing systems,\footnote{Similar to
the benchmarking of classical computers, with the LINPACK benchmarks
\cite{petitet_whaley_dongarra_cleary, doi:10.1002/cpe.728} being used to build
the TOP500 ranking of supercomputers \cite{Dongarra2011}.} and tracking the
performance of a given system over time, provide little information on how
different combinations of the stack's components could change system
performance. For this reason, holistic benchmarking should, as much as
possible, make explicit the variable components of the stack, and
systematically vary them to identify how a particular component affects
system-level performance. This ensures these benchmarks identify improvements
due to changes at all layers of the \stack{}, towards which, for example,
improvements in the compilation strategy can contribute significantly.

The benchmarks defined here have two parts: a \emph{circuit class} and a
\emph{figure of merit}. The circuit class describes the type of circuit to be
run by the system, and the figure of merit quantifies performance of the system
when running circuits from that class.

Quantum computing systems are used for particular applications, so the circuit
classes should test the performance of a system in those arenas
\cite{Linke3305}. Assessing the performance of \stack{}s when implementing
circuits derived from applications ensures that the results are stronger
indicators of whether those applications can be successfully implemented than
they would otherwise be (for example, by numerical simulation of the
application using a noise model).

At least two notions have been put forth for defining such classes. The first
is benchmarks based on often-used quantum algorithmic primitives
\cite{blume2019volumetric}, the circuit classes being primitives of Grover
iterations and Trotterized Hamiltonian simulation. This approach has the
advantage that such computational primitives will be ubiquitous in quantum
applications, but has the disadvantage that implementations on near-term
quantum computers may yield poor performance.

The second is for the circuit class to be a particular instance of an
application. Such benchmarks have been defined in the context of quantum
simulation \cite{mcardle2018quantum, McCaskey2019, dallaire2020application,
PhysRevLett.120.210501, arute2020hartreefock}, quantum machine learning
\cite{Dunjko_2018, Linke3305, Benedetti2019, hamilton2019error,
PhysRevA.99.062323}, discrete optimisation \cite{arute2020quantum,
willsch2019benchmarking, bengtsson2019quantum, pagano2019quantum}, and
\supremacy{} \cite{preskill2012quantum, harrow2017quantum, Arute2019,
boixo2018characterizing}. This approach has the advantage that the definition
of success when running a circuit from the class is fairly straightforward, but
with the disadvantage that performance, as measured by one instance of an
application, may not be predictive of performance for the application
generically, or another application. Further, while the wide selection of
circuits presented in the aforementioned literature covers an array of
applications, deriving benchmarks independently of each other may result in a
lack of coverage, or unnecessary repeated coverage, of circuit classes.

The ``application-motivated'' circuit classes defined in this work draw
inspiration from \cite{blume2019volumetric} (looking at computational
primitives) but also from the literature above (focusing on computational
primitives of near-term quantum computing applications). This approach has the
advantage that a system which does well on an application-motivated benchmark
should do well in running the application the benchmark was derived from.
Further, by considering application primitives, the result of the benchmarks
apply to the application generally, and not only to a particular instance of
the application.

We introduce 3 such ``application-motivated'' circuit classes, which cover
varying depth regimes and are (somewhat) controllable in depth. The definitions
and motivations for each class are given in \secref{sec:circuit classes} but,
in brief, the classes -- as labelled by their depth regimes -- and the
applications that motivate them, are:
\begin{description}
    \item[Shallow:] Inspired by hardware-efficient ansatze
        \cite{sukin2019expressibility, Kandala2017} which may be useful for
        near-term quantum machine learning and chemistry applications
        \cite{coyle2019born, du2018expressive, hubregtsen2020evaluation}. The
        width of \shallow{} grow with a minimal increase in depth, allowing the
        impact of including many qubits to be explored.
    \item[Square:] Inspired by the circuits used to calculate a system's
        quantum volume \cite{cross2018validating}. These circuits utilise gates
        sampled uniformly at random from all $\mathrm{SU}\brac{4}$ gates,
        making them a test of general-purpose, programmable quantum computers.
    \item[Deep:] Inspired by product formula circuits, including state
        preparation circuits used in the variational quantum eigensolver (VQE)
        algorithm\cite{Berry2007, peruzzo2014variational, Romero_2018}. The
        depth of these circuits grows most quickly with width, giving a
        thorough coverage of depth regimes.
\end{description}
Importantly, as discussed in \secref{sec:circuit classes}, these circuits are
complementary; they cover a wide selection of applications and circuit types.
By covering many applications with few circuit classes, we are able to
concisely present far- reaching results, and minimise the number of
computations to be performed in deriving those results.

How well a stack executes a circuit is assessed here via continuous figures of
merit, rather than binary ones which may only verify correctness. This is
because the outcomes from noisy devices will likely not be correct, while
information about closeness to the correct answer is still highly valuable.
Further, techniques for the verification of universal quantum computation
requires many qubits, or qubit communication, or both, none of which are
accessible using currently-available devices \cite{Gheorghiu2019,
mahadev2018classical}. 

We use three figures of merit, calculated using classical computers: heavy
output generation probability \cite{aaronson2016complexity}, cross-entropy
difference \cite{boixo2018characterizing}, and \lone{}.  Estimating each of
these figures of merit requires knowledge about the \emph{ideal} (noise-free)
outcome (bitstring) probabilities.

In practice, calculating them requires direct simulation of the circuit under
consideration. Consequently, scaling to tens or hundreds of qubits will be
challenging in general, particularly if the \lone{} is used as the figure of
merit. However, by considering circuits with a few qubits, the circuits can be
simulated classically, and insight into the behaviour of larger devices
\cite{Wiebe_2015, Arute2019} can be gained.

However, significant improvements in the time needed to perform benchmarks can
be made if the circuits and figures of merit are developed jointly. Indeed, the
calculation of heavy output generation and cross-entropy difference requires
only a polynomial number of samples from a quantum device for some circuit
classes. Importantly, we show that this is the case for \deep{} and \square{}.
Therefore, \deep{} provide the first instance of chemistry-motivated circuits,
which are likely not possible to classically simulate in general, but which can
be benchmarked using polynomially many samples from the output distribution
from a device implementing them.

We refer to a set of benchmarks as a \emph{benchmarking suite}, each benchmark
being defined by unique combinations of each circuit class and figure of merit.
Using a benchmarking suite enables the derivation of broad insights about the
behaviour and performance of a system across a wide variety of possible
applications. Their varying demands on quantum computing resources (e.g.,
qubits, achievable depth) allows for the exploration of the best routes to
extract the most utility from near-term quantum computers.  In sum, in order to
predict the systems' performance in practice, our benchmarking approach is both
application-motivated, holistic and full \stack{}. For the reasons stated
above, we regard all of these properties to be necessary of a standard
benchmark suite.

Here, we will benchmark systems made available by IBM Quantum using the
benchmarking suite defined in this work, and investigate two components of the
stack: the \emph{compilation strategy} used to map an abstract circuit onto one
that is executable on a quantum computer, and the \emph{device} used to run the
compiled circuit and return the results.\footnote{While the particular
systems used here have other components (such as pulse synthesisers), we do not
look at the impact of those pieces on full-stack performance.}

The remainder of this paper is comprised as follows: \secref{sec:circuit
classes} details the circuit classes, including algorithms for generating the
circuits; \secref{sec:metrics} explains the figures of merit we use;
\secref{sec:stack} introduces the software stack, as well as hardware made
available by IBM Quantum, that comprise the systems we'll be benchmarking; and
\secref{sec:results} shows the results of our benchmarking. We conclude in
\secref{sec:conclusion}.

\section{Figures of Merit}
\label{sec:metrics}

Suppose a quantum computer runs an $n$ qubit circuit $C$, on the input
$\ket{0}^{n}$. Repeated runs produce a set of classical bitstrings $x_1, ... ,
x_k$ (with $k$ being the total number of runs). Figures of merit compare $p_C
\brac{x_j} = \mod{\bra{x} C \ket{0^n}}^2$, the ideal output probabilities of
each $x_j$ in $C$, and $\dist{D}_C \brac{x_j}$, the probability that $x_i$ is
produced by a real implementation, which may be noisy.

In practice, direct access to $\dist{D}_C \brac{x_j}$ is impossible, because
the number of bitstrings generally grows exponentially with circuit width.
Consequently, for a fixed $k$, there will be some bitstrings which are not
observed. The $k$ bitstrings which are observed can be post-processed to
\emph{estimate} a given figure of merit. Statistical uncertainty in the value
of figures of merit can arise in this way.  On the other hand, simulations can
be used to calculate $p_C \brac{x_j}$ directly, and the figures of merit we
discuss will make use of this.

As such, focusing on figures of merit which can be reasonably approximated from
samples is important. This is to say with a number of samples from $\dist{D}_C$
which is polynomial in the size of the circuit, as opposed to the exponential
number which would be required for a full characterisation. This focus allows
for a better scaling in the resource requirements as the size of the circuit
grows.

The circuits we investigate are small enough that $\dist{D}_C$ can be well
characterised by a reasonable absolute number of samples. As such we will also
include a discussion and comparison with the \lone{}, motivated by its relation
to several theoretical results, notably \thmref{thm:iqp lone}.\footnote{In the
case that sufficient samples can be obtained to characterise $\dist{D}_C$ well
, several other figures of merit are accessible, such as the KL-divergence. We
limit our investigations to the three discussed to remain thorough but concise,
but invite further analysis of the data generated by our experiments. This data
can be found under `Data Availability' towards the end of the paper.}

The remainder of this section outlines three figures of merit: Heavy output
probability, Cross-entropy difference, and \lone{}. We detail: their
definition, the continuous range of values they can take, their dependence on
noise, and the procedure for calculating their value from samples produced by
an implementation. These figures of merit have been developed in other works,
and so in this section we explore these properties in so far as they facilitate
their use and justify our choosing them.


\subsection{Heavy Output Probability}
\label{sec:hog}

Heavy Output Generation \cite{aaronson2016complexity} (HOG) is the problem of
producing bitstrings that are predominantly those that are the most likely in
the output distribution of $C$. That is to say, outputs with the highest
probability in the ideal distribution should be produced most regularly.

If the ideal distribution is sufficiently far from uniform, HOG provides a
means to distinguish between samples from the ideal distribution and a trivial
attempt to mimic such a sampling procedure; namely by producing uniformly
random bitstrings. Although a simply stated problem, which is verifiable by a
classical device using a polynomial number of samples from $\dist{D}_C$, this
task is conjectured to be hard for a classical computer to perform in general
\cite{aaronson2016complexity}.


\defpar{}

An output $z \in \curlbrac{0,1}^n$ is \emph{heavy} for a quantum circuit $C$,
if $p_C\brac{z}$ is greater than the median of the set $\curlbrac{p_C \brac{x}
: x \in \curlbrac{0,1}^n}$.  We can define the probability that samples drawn
from a distribution $\dist{D}_C$ will be heavy outputs in the distribution
$p_C$, called the \emph{heavy output generation probability of $\dist{D}_C$},
as follows:
\begin{equation}
    \hog{\dist{D}_C}{p_C} = \sum_{x \in \curlbrac{0 , 1}^n} \dist{D}_C \brac{x}
    \delta_C \brac{x}.
\end{equation}
Here $\delta_C \brac{x} = 1$ if $x$ is heavy for $C$, and $0$ otherwise.

In the case $\hog{p_C}{p_C} \approx 1/2$, outputs which are heavy have similar
probabilities of occurring as outputs that are not. Such distributions are well
approximated by the uniform distribution.  Hence, for $\hog{\dist{D}_C}{p_C}$
to help us distinguish between an ideal implementation of $C$ and a trivial
attempt to mimic it by generating random bitstrings, $\hog{p_C}{p_C}$ should be
greater than $1/2$. In fact, $\hog{p_C}{p_C}$ is expected to be $(1 +
\log{2})/2 \approx 0.846574$ \cite{aaronson2016complexity} for circuit classes
whose distribution of measurement probabilities, $p$, is of the exponential
form $\mathrm{Pr} \brac{p} = Ne^{-Np}$, where $N=2^n$.\footnote{This is also
commonly referred to as the Porter-Thomas distribution \cite{porterthomas}.}
This is discussed at length in \appref{app:far from uniform} where we will
demonstrate that the \deep{} and \square{} classes have this property. When the
output distributions of a class of circuits is shown to take this form it is
meaningful to define the Heavy Output Generation problem.
\begin{problem}[Heavy Output Generation \cite{aaronson2016complexity}]
    \label{prob:hog}
    Given a measure $\mu$ over a class of circuits, the family of distributions
    $\curlbrac{\dist{D}_C}$ is said to satisfy HOG if the following is true.
    \begin{equation}
        \mathbb{E}_{C \leftarrow \mu} \sqrbrac{ \hog{\dist{D}_C}{p_C} } \geq
        \frac{2}{3} 
    \end{equation}
\end{problem}
The classical harness of HOG can be related to the difficulty of calculating
the output probabilities of quantum circuits \cite{aaronson2016complexity}.  In
particular, such calculations have been argued to be beyond classical computers
using polynomial resources in the case of circuits similar to the \square{} of
\secref{sec:random circuits} \cite{aaronson2016complexity}.

For the circuit classes in \secref{sec:circuit classes} which have
exponentially distributed output probabilities, we will often consider the
largest $n$ for which distributions $\curlbrac{\dist{D}_{C_n}}$ solve
\probref{prob:hog}. This approach is inspired by quantum volume and is useful
as an indicator of the largest Hilbert space accessible to a \stack{}
\cite{cross2018validating}.  For circuit classes with output probabilities that
are not distributed as such, we will explicitly calculate the ideal heavy
output probability as a point of comparison.


\classver{}

We approximate $\hog{\dist{D}_C}{p_C}$ in a number of operations which grows
exponentially with the number of qubits, but using only a polynomial number of
samples from the real distribution $\dist{D}_C$. To do so we evaluate the
following expression, where the ideal probabilities $p_C \brac{x}$ must be
calculate in order to determine $\delta_C \brac{x}$, and where $x_1, ... , x_k$
are samples drawn from $\dist{D}_C$.
\begin{equation}
    \frac{1}{k} \sum_{i = 1 , ... , k} \delta_C \brac{x_i}
\end{equation}
By the law of large numbers, this converges to $\hog{\dist{D}_C}{p_C}$ in the
limit of increasing sample size. 


\quantimp{}

In the case of extreme noise, and the convergence of the real distribution
$\dist{D}_C$ to the uniform distribution $\dist{U}$, $\hog{\dist{D}_C}{p_C}
=1/2$. This is compared to the case where $\dist{D}_C = p_{C}$, when we would
expect to have $\hog{\dist{D}_C}{p_C} = (1 + \log{2})/2$. 

Accordingly, in the results of \secref{sec:results}, we say a \stack{} has
performed well if, on average over the class of circuits,
$\hog{\dist{D}_C}{p_C}$ is between $\brac{1 + \log{2}}/2$ and $2/3$, with
$\brac{1 + \log{2}}/2$ being best of all. A poorer performance is indicated by
average values between $2/3$ and $1/2$, with $1/2$ being worst of all. This
gives a continuum of values, with an intuitive interpretation, with which to
compare \stack{}s, which we call \HOG{}.


\subsection{Cross-Entropy Difference}
\label{sec:xed}

Cross-entropy benchmarking \cite{boixo2018characterizing} relates to the
average probability, in the ideal distribution, $p_C$, of the outputs which are
sampled from the empirical distribution, $\dist{D}_C$. For distributions which
are far from uniform, this measure can be used to distinguish an ideal from a
noisy implementation. Ideal implementations will regularly produce the higher
probability outputs, obtaining a high benchmark value, while even a small shift
in the distribution will lower the value.

The cross-entropy difference can be calculated using exponential classical
resources, from a polynomial number of samples from a quantum computer, which
allows for its utilisation in benchmarking smaller quantum devices
\cite{neill2018blueprint, boixo2017fourier, boixo2018characterizing,
Arute2019}. There are also well-developed means by which this quantity can be
used as a means of extrapolating from the behaviour of smaller devices to that
of larger devices, which might demonstrate \supremacy{} \cite{Arute2019}.


\defpar{}

The \emph{cross-entropy} captures one's surprise when sampling from $\dist{D}$
when one is expecting $\dist{D}'$:
\begin{definition}[Cross-Entropy]
  \label{def:xe}
  \xedef{}
\end{definition}
The \emph{cross-entropy difference} $\mathrm{CED} \brac{\dist{D} , \dist{D'}}$
is $\mathrm{CE} \brac{\dist{U} , \dist{D'}} - \mathrm{CE} \brac{\dist{D} ,
\dist{D'}}$, where $\dist{U}$ is the uniform distribution.  Hence, the
cross-entropy difference can be seen as answering ``Is the distribution
$\dist{D}'$ best predicted by $\dist{D}$ or by the uniform distribution?''. 

The comparison to the uniform distribution which the cross-entropy difference
provides is valuable as, if an honest attempt is being made to recreate a
distribution, at worst $\dist{U}$ could be produced.

In some cases, the cross-entropy also gives an estimate for the average circuit
fidelity \cite{boixo2018characterizing}, facilitating the characterisation of
noise levels in implementations of quantum circuits. While \XE{} on its own
cannot be used to distinguish error channels, in combination with the
techniques introduced here, it can provide insight into this information.


\classver{}

By the law of large numbers, the following expression converges to $\mathrm{CE}
\brac{\dist{D}_{C}, p_C}$, where $x_1 , ... , x_k$ are samples drawn from
$\dist{D}_{C}$.
\begin{equation}
  \label{eq:classical xe}
  \frac{1}{k} \sum_{i = 1 , ... , k} \log \brac{\frac{1}{p_C \brac{x_i}}}
\end{equation}
This can be used by a classical computer to approximate the value for
$\mathrm{CED} \brac{\dist{D}_{C}, p_C}$.  While only a polynomial number of
samples $x_i$ are required, the best known algorithms for calculating $p_{C}
\brac{x_i}$ for an arbitrary circuit $C$ require exponential time or space on a
classical computer. It should be expected, therefore, that calculating the
cross-entropy difference consumes exponential classical compute resources in
general.

In our case, to avoid requiring the inverse of $0$ in this approximation, we
chose to use 
\begin{equation}
    \max \curlbrac{ p_C \brac{x} , 2^{-n^{2}}}.
\end{equation}
This choice of an inverse exponential in the number of qubits is inspired by,
although not directly derived from, the average case supremacy results related
to random circuits \cite{bouland2018quantum, movassagh2018efficient}.


\quantimp{}

$\mathrm{CE} \brac{\dist{D}_{C},p_{C}}$, reduces to the entropy, $H
\brac{p_{C}}$, of $p_{C}$ when there is no noise, as in this case $p_C =
\dist{D}_C$. Further, when the probabilities $p_C \brac{x}$ are approximately
independent and identically distributed according to the exponential
distribution, $H \brac{p_{C}} = \log 2^n + \gamma -1$
\cite{boixo2018characterizing}, where $\gamma$ is Euler's constant.

If $\dist{D}_{C} \brac{x}$ is uncorrelated with $p_{C} \brac{x}$, the following
result holds \cite{boixo2018characterizing}:
\begin{equation}
    \mathbb{E}_C \sqrbrac{\mathrm{CE} \brac{\dist{D}_C , p_C}} = \log 2^n + \gamma
\end{equation}
$\dist{D}_{C} \brac{x}$ and $p_{C} \brac{x}$ are uncorrelated if, for example,
$\dist{D}_{C}$ is the uniform distribution, or, in the case of demonstrations
of \supremacy{}, if $\dist{D}_C$ is the output of a polynomial cost classical
algorithm \cite{boixo2018characterizing}.

These results allow us to identify the extreme values taken by the
cross-entropy difference:
\begin{align}
  \dist{D}_{C} = p_{C} &: & \mathrm{CED} \brac{\dist{D}_{C}, p_{C}} & = 1 \\
  \dist{D}_{C} = \dist{U} &: & \mathrm{CED} \brac{\dist{D}_{C}, p_{C}} & =
  0
\end{align}
As such, the cross-entropy difference gives a value between $0$ and $1$ which
measures the accuracy of the implementation of a circuit, the calculation of
which is called \XE{}. 

Hence, in the results of \secref{sec:results} we will say that better
performance is indicated by a value, averaged over the class of circuits,
closer to $1$, and worse by a value closer to $0$. Notice that values above 1,
or below 0, are possible for individual circuits if, for example, the ideal
output distribution happens to be very skewed towards heavy outputs, or if
unlikely outcomes occur very often, respectively. Unlike in the case of \HOG{},
we do not give a value which a score above would indicate good
performance.\footnote{Where such a value has been defined, it is perhaps
surprisingly close to $0$, and so the uniform distribution \cite{Arute2019}.}


\subsection{\titlecap{\lone{}}}
\label{sec:l1 norm}

The \lone{} between two probability distributions measures the total difference
between the probabilities the distributions assign to elements of their sample
space. Such a metric is sufficiently strong that for several classes of quantum
circuits it is known that classical simulation of all circuits in the class to
within some \lone{} of the ideal distribution would contradict commonly held
computational complexity theoretic conjectures \cite{aaronson2011computational,
bremner2016average, bouland2018quantum}.

Unlike the previous two figures of merit, approximating the \lone{} requires a
full characterisation of the ideal output distribution. In the cases where few
qubits are considered, as is so here, it is possible to perform such
characterisations. For larger qubit counts, the cross-entropy benchmarking and
heavy output generation are the preferred benchmarking schemes.


\defpar{}

In the case of distributions over the sample space $\curlbrac{0,1}^n$, the
\lone{} is defined as follows.
\begin{definition}[\lone{}]
    For distributions $\dist{D}$ and $\dist{D}'$ over the sample space
    $\curlbrac{0,1}^n$ the \lone{} between them is
    \begin{equation}
        \label{eq:lone}
        \ell_1 \brac{\dist{D},\dist{D}'} = \sum_{x \in \curlbrac{0,1}^{n}}
        \mod{\dist{D} \brac{x} - \dist{D}' \brac{x}}.
    \end{equation}
\end{definition}
Because of its independence from probability values themselves, it is regarded
as reasonable to require quantum computers to produce samples from
distributions within some \lone{} of the ideal distribution. This is as opposed
to measures of distance, such as multiplicative error, which require zero
probability outcomes are preserved in the presence of noise, but for which very
strong connections to \supremacy{} also exist \cite{bremner2011classical}.


\classver{}

In this work we will approximate the \lone{} between the ideal and real
distributions using samples from the real distribution. Given samples $s =
\curlbrac{x_1 , ... , x_m}$ from $\dist{D}_C$, let $s^x$ be the number of times
$x$ appears in $s$. Define $\widetilde{\dist{D}_C}$ by $\widetilde{\dist{D}_C}
\brac{x} = (s^x)/m$. Then the approximation we will use for $\ell_1
\brac{\dist{D}_C,p_C}$ is $\ell_1 \brac{\widetilde{\dist{D}_C}, p_C}$.


\quantimp{}

An ideal implementation of a circuit would result in $\ell_1 \brac{\dist{D}_C,
p_C} = 0$. As such, in the results of \secref{sec:results} we will say that a
\stack{} performs better the closer this value is to $0$. However, noise will
likely make it incredibly difficult for even fault tolerant quantum computers
to achieve a \lone{} of $0$.  Hence bounds, such as that discussed in
\thmref{thm:iqp lone}, are often put on the value instead. For the circuit
classes for which such results are relevant, as we discuss in
\secref{sec:circuit classes}, we will use these results to define a values
below which a performance can be considered as good.  Once again, the \lone{}
takes a continuous range of values allowing for comparison between
implementations of circuits.


\subsection{Metric Comparison}
\label{sec:metric comparison}

Unfortunately, \XE{} and \HOG{} cannot be used to bound the \lone{}
\cite{bouland2018quantum}, which, as noted in \secref{sec:l1 norm}, provides
strong guarantees of demonstrations of \supremacy{}.\footnote{Interestingly,
empirical result show a slight negative correlation between \lone{} and
experimental heavy output probability (normalized to the heavy output
probability of an ideal device). This relationship is discussed in
\appref{app:l1_hog}, which provides empirical details of this relationship.}
That is, the \lone{} provides uniquely (amongst the metrics studied here)
strong assurances about \supremacy{}.  In general, such assurances come at the
cost of requiring classical compute resources which grow exponentially with the
circuit size.

In the case of \HOG{} and \XE{} only polynomially many single output
probabilities are required, allowing the utilisation of \emph{Feynman
simulators} \cite{aaronson2016complexity}. These compute output bitstring
amplitudes by adding all Feynman path contributions. This extends the domain of
classical simulation by overcoming the memory storage problem, establishing the
frontier of what's possible on classical computers \cite{Villalonga2019,
villalonga2019establishing, Arute2019}. However, this method still requires
exponential time to perform and so reaches its own limit for large numbers of
qubits.

Importantly, since $\hog{\dist{D}_C}{p_C}$ and $\mathrm{CE} \brac{\dist{D}_C,
p_C}$ are expectations of different functions of ideal output probabilities,
$\delta \brac{p_C}$ and $-\log \brac{p_C}$ respectively, over the experimental
output distribution, they capture different features of the outputs
\cite{bouland2018quantum}.  In fact $\hog{\dist{D}_C}{p_C}$ can also be used to
approximate circuit fidelity, however the standard deviation of the estimator
is larger than that for $\mathrm{CE} \brac{\dist{D}_C, p_C}$ \cite{Arute2019}.

\section{Circuit Classes}
\label{sec:circuit classes}

This section presents the formal definitions of the circuits used in this work,
while also identifying the motivations for their use in benchmarking. These
motivations include both the class of applications they represent and the
properties of the \stack{}s that they will probe. Collectively, this selection
of circuit classes encompass an array of potential applications of quantum
computing, covering circuits of varied depth, connectivity, and gate types.


\subsection{\Shallow{}: IQP}
\label{sec:shallow}

Instantaneous Quantum Polytime (IQP) circuits \cite{iqporig} consist of
commuting gates and, as well as being simpler to implement than universal
quantum circuits, are believed, even in the presence of noise, to be impossible
to simulate efficiently using classical computers \cite{bremner2011classical,
bremner2016average, Bremner2017achievingquantum}. This has allowed for the
application of noisy quantum technology in areas such as machine learning
\cite{coyle2019born, du2018expressive, Havl_ek_2019} and interactive two-player
games \cite{mills2017information, iqporig}.  The \emph{shallow} class of
circuits introduced here, whose depth increases slowly with width, is a
subclass of IQP circuits. 


\defpar{}

An $n$-qubit IQP circuit consists of gates that are diagonal in the Pauli-\X{}
basis, acting on the $\ket{0}^n$ state, with measurement taking place in the
computational basis. For this class of circuits, \thmref{thm:iqp lone} applies.
\begin{theorem}[Informal \cite{bremner2016average}]
    \label{thm:iqp lone} 
    Assuming either one of two conjectures, relating to the hardness of
    approximating the Ising partition function and the gap of degree 3
    polynomials, and the stability of the Polynomial Hierarchy,\footnote{The
    non-collapse of the Polynomial Hierarchy is widely conjectured to be true
    \cite{10.1145/800141.804678}.} it is impossible to classically sample from
    the output probability distribution of any IQP circuit in polynomial time,
    up to an \lone{} of $1/192$.
\end{theorem}

This class is called ``instantaneous'' because these gates commute with one
another, which has the benefit of reducing the amount of time that the quantum
state will need to be stored. Further, the impossibility of simulating IQP
circuits is shown to hold when restricted by physically motivated constraints
such as limited connectivity and constant error rates on each qubit
\cite{Bremner2017achievingquantum}. This makes the class an exciting one to
explore with near-term technology.

An equivalent, commonly-considered definition is that IQP circuits consist of
gates diagonal in the Pauli-\Z{} basis, sandwiched between two layers of
Hadamard gates acting on all qubits. \algref{alg:iqp circuits} is used to
generate IQP circuits of this form. Note that \algref{alg:iqp circuits} limits
the connectivity allowed between the qubits, so it does not generate \emph{all}
circuits in the IQP class.

\begin{figure}
\begin{algorithm}[H]
    \caption{The pattern for building \shallow{}.}
    \label{alg:iqp circuits}
    \inout{Number of qubits, $n \in \mathbb{Z}$}{$7$}{Circuit, $C_n$}
    \begin{algorithmic}[1]
        \State Initialise $n$ qubits, labelled $q_1 , ... , q_n$, in the
          state $\ket{0}$.
        \State
        \ForAll{$i \in \curlbrac{1, ... , n}$}
            \State Act \H{} on $q_i$
        \EndFor
        \State
        \State Let $G_n$ be the graph indicating between which qubits a \CZ{}
            gate can act. Let $G_n$ in this case be a random binomial graph,
            $G_n$, with $n$ vertices and edge probability $0.5$, post selecting
            on those that are connected and have degree less than 4.
        \State
        \ForAll{edges $\curlbrac{i,j}$ in $G_n$}
            \State Act \CZ{} between $q_i$ and $q_j$
        \EndFor
        \State
        \ForAll{$i \in \curlbrac{1, ... , n}$}
            \State Generate $\alpha_i \in \sqrbrac{0, 2 \pi}$ uniformly at
                random.
            \State Act $\RZ{} \brac{\alpha_i}$ on $q_i$ .
        \EndFor
        \State
        \ForAll{$i \in \curlbrac{1, ... , n}$}
            \State Act \H{} on $q_i$
        \EndFor
        \State
        \State Measure $q_1, ...  , q_n$ in the computational basis
    \end{algorithmic}
\end{algorithm}
\end{figure}

The depth of this circuit may be arrived at by observing that finding a valid
order in which to apply the \CZ{} gates in the circuit is equivalent to finding
an edge colouring of the graph $G_n$. By Vizing's theorem, it is known that a
colouring of undirected graphs using at most one more colour than the maximum
vertex degree exists.  Constructive proofs of Vizing's theorem therefore
demonstrate that a $4$-colouring can be found in polynomial time
\cite{MISRA1992131}.\footnote{It is possible that the chromatic index of a
particular graph is lower than $4$, either because: the maximum degree is less
than $3$, as could be the case with these random graphs; or because a
$3$-colouring exists, as is the lower bound on the chromatic index of an
undirected graph of maximum vertex degree $3$.  However a $4$-colouring
certainly provide an upper bound and so it suffices for our discussion as we
are concerned with upper bounding the depth of the circuit.} \algref{alg:iqp
circuits} includes discrete randomness over the graphs, $G_n$, and continuous
randomness over the rotation angles, $\alpha_i$.

The close connection, through \thmref{thm:iqp lone}, of \supremacy{} and
\shallow{},\footnote{While \thmref{thm:iqp lone} is a worst case hardness
result, and may not apply to \shallow{}, we regard their performance as
indicative of that of those for which it does. Indeed, similar hardness results
to \thmref{thm:iqp lone} exist for other families of sparse, constant depth IQP
circuits \cite{bremejo2018architecture}.} explicitly measured in \lone{},
provides a measure of a \stack{}'s quality. This is namely, by analysing the
closeness of the distributions it produces to the ideal ones, as measured by
the \lone{}, and comparing this value to $1/192$.

However, in the case of \shallow{}, the output probabilities are not
exponentially distributed. As expanded on in \secref{sec:xed} and
\secref{sec:hog}, this property allows us to simplify calculations required
when performing both \XE{} and \HOG{}. In particular the theoretical value of
heavy output probability for circuits with exponentially distributed output
probabilities cannot be used here.

Instead, we use the empirical value of the ideal heavy output probability, in
the place of a theoretically derived one, as a point of comparison with the
behaviour of the \stack{} being benchmarked. This approach requires calculation
of all output probabilities and summation of the probabilities of those that
are heavy, which is to say have greater than the median ideal probability.
This can be done for the small circuits investigated here, but allows for the
benchmarking of fewer qubits than would be accessible if a theoretical value
was known.

\motpar{}

The design of circuits in \algref{alg:iqp circuits} may be compared to other
sparse IQP circuits \cite{Bremner2017achievingquantum}, IQP circuits on 2D
lattices \cite{Bremner2017achievingquantum, bremejo2018architecture}, and
random 3-regular graphs used for benchmarking \cite{arute2020quantum}.  In our
case we aim to avoid favouring particular architectures, and so avoid 2D
lattices. In addition we avoid highly structured 3-regular graphs, in favour of
allowing reduced vertex connectivity.\footnote{As $n$ grows the probability
that the maximum degree is at most 3 will tend to zero, and so \algref{alg:iqp
circuits} may become infeasible to implement. While for the number of qubits
considered in this paper \algref{alg:iqp circuits} works well, for larger $n$
it may become necessary to alter the graph generation step, possibly to the use
of $3$-regular graphs, for which the fast random generation algorithms have
been extensively researched and developed. However, regular graphs are too few
in number for graphs with low numbers of qubits to provide a good benchmark set
of many possible circuits, and so we avoid them here.} This avoids preferring
very particular applications of low depth IQP circuits; instead exploring a
variety of applications simultaneously. It is also of interest that there are
sparse IQP circuits for which verification schemes exist \cite{Hangleiter_2017,
bremejo2018architecture} although the connectivity is too architecture-specific
for our purposes, with the verification scheme requiring limits to the
measurement noise which we cannot guarantee.

Before compilation, \shallow{} have constant depth, allowing us to measure the
impact of increasing circuit width independently of increasing circuit depth.
Further, because \algref{alg:iqp circuits} limits the connectivity allowed
between the qubits, the increase in circuit depth due to compilation onto
limited-connectivity architectures is also minimised, while avoiding a choice
of connectivity favouring one device in particular. By bounding connectivity,
but allowing all connections in principle, we avoid biasing against
architectures that allow all-to-all connectivity, which would still perform
well.

In summary, inspired by a variety of near-term applications of IQP circuits
\cite{coyle2019born, du2018expressive, mills2017information, iqporig,
Havl_ek_2019}, we introduce \shallow{} in \algref{alg:iqp circuits}.
Performance when implementing these circuits is indicative of the performance
when implementing those applications, but also, more generally, of applications
requiring circuits which grow slowly in depth.  As a result, these circuits
also probe the performance of a \stack{} in fine-grained detail by measuring
the impact of including more qubits (quasi-) independently of increasing
circuit depth. This is useful for understanding the performance of a device
being utilised for applications whose qubit requirement grows more quickly than
the circuit depth.


\subsection{\Square{}: Random Circuit Sampling}
\label{sec:random circuits}

While circuits required for applications are typically not random, sampling
from the output distributions of random circuits built from two-qubit gates has
been suggested as a means to demonstrate \supremacy{} \cite{bouland2018quantum,
movassagh2018efficient, boixo2018characterizing, aaronson2016complexity}.  By
utilising uniformly random two-qubit unitaries and all-to-all connectivity,
\emph{\square{}}, introduced now, also provide a benchmark at all layers of the
\stack{}. With a circuit depth that grows linearly with the number of qubits,
\square{} also probes the performance of a \stack{} in ways not accessed by
\shallow{}.


\defpar{}

A random circuit, for a fixed number of qubits $n$ and graph $G_n$, is
generated by applying $m = \poly{n}$ uniformly random two-qubit $\mathrm{SU}
\brac{4}$ gates between qubits connected by edges of $G_n$. Here, ``uniformly
random'' means according to the Haar measure.

\emph{Random Circuit Sampling} (RCS) is the task of producing samples from the
output distribution of random circuits. To perform RCS approximately is to
sample from a distribution close to that produced by the random circuit. For
certain classes of graphs $G_n$, this task has been shown to be hard in the
average case \cite{bouland2018quantum, movassagh2018efficient}, as outlined in
\thmref{thm:average case rc hardness}. 

\begin{theorem}[Informal \cite{bouland2018quantum}]
    \label{thm:average case rc hardness}
    There exists a collection of graphs $G_n$, with one for each $n$, and
    procedure for generating random circuits respecting each $G_n$, for which
    there is no classical randomised algorithm that performs approximate RCS,
    to within inverse polynomial \lone{} error, for a constant fraction of the
    random circuits.
\end{theorem}

The conditions imposed on which graphs and circuit generation procedures are
covered by this theorem are quite mild, but in particular this can be done
using circuits with depth $\mathcal{O}\brac{n}$ acting on a 2D square lattice
\cite{bouland2018quantum, aaronson2016complexity}.  Note that in that case
$G_n$ are not generated at random, but systematically according to $n$, unlike
the graphs used to generate \shallow{} in \algref{alg:iqp circuits}. While this
is relevant for devices built using superconducting technology
\cite{Arute2019}, we wish to avoid biasing in favour of this technology in
particular.

The circuits used here -- which are almost identical to those used for the
quantum volume benchmark \cite{cross2018validating} -- are generated according
to \algref{alg:random circuits}. We refer to this class of circuits as
\square{}, and note that they consist of $n$ \emph{layers} of two-qubit gates
acting between a bipartition of the qubits.  There is discrete randomness over
the possible bipartition of the qubits, and continuous randomness over the
matrix elements of two-qubit $\mathrm{SU} \brac{4}$ gates selected uniformly at
random from the Haar measure.

\begin{figure}
\begin{algorithm}[H]
    \caption{The pattern for building \square{}.}
    \label{alg:random circuits}
    \inout{Number of qubits, $n \in \mathbb{Z}$}{$n$}{Circuit, $C_n$}
    \begin{algorithmic}[1]
        \State Initialise n qubits, labelled $q_1 , ... , q_n$, in the
            state $\ket{0}$
        \State
        \For{each layer t up to depth n}
            \State \Comment The contents of this for loop constitutes a
                \emph{layer}. The choice of the number of layers used here is
                discussed in \appref{app:far from uniform square}.
            \State
            \State Divide the qubits into $\floor{\frac{n}{2}}$ pairs
                $\curlbrac{q_{i,1}, q_{i,2}}$ at random.
            \ForAll{$i \in \mathbb{Z}$, $0 \leq i \leq \floor{\frac{n}{2}}$}
                \State Generate $\U{}_{i,t} \in \mathrm{SU} \brac{4}$ uniformly
                    at random according to the Haar measure.
                \State Act $\U{}_{i,t}$ on qubits $q_{i, 1}$ and $q_{i, 2}$.
            \EndFor
        \EndFor
        \State
        \State Measure all qubits in the computational basis.
    \end{algorithmic}
\end{algorithm}
\end{figure}

In addition, \square{} fulfil the necessary conditions to apply HOG, as defined
in \probref{prob:hog}. Namely, the distribution $p_C$ is sufficiently far from
uniform in the required sense, as introduced in \secref{sec:hog}, which we
demonstrate in \appref{app:far from uniform square}.


\motpar{}

By utilising uniformly random two-qubit unitaries, \square{} provide a
benchmark at all layers of the \stack{}. In particular this tests the ability
of the device to implement a universal gate set, the diversity and quality of
the gates available, and the compilation strategy's ability to decompose these
gates to the native architecture. Further, as quantum circuits can always be
approximated up to arbitrary precision using two-qubit unitary gates
\cite{nielsen2002quantum}, \square{} can help us understand the performance of
\stack{}s when implementing computations requiring a universal gate set, which
is less the case for \shallow{} and \deep{}.

By allowing two-qubit gates to act between any pair of qubits in the uncompiled
circuit, \square{} avoid favouring any device in particular \cite{Arute2019,
boixo2018characterizing, aaronson2016complexity}. This choice adheres closely
to our motivations of being hardware-agnostic. In addition, assuming all-to-all
connectivity passes the burden of mapping the circuit onto the device to the
compilation strategy, which is in line with our wish to benchmark the full
\stack{}. That said, any architecture whose coupling map closely mirrors the
uncompiled circuit will be advantaged, as even a naive compilation strategy
will perform well in that case.

In \cite{cross2018validating} similar circuits are used but with all-to-all
connectivity restricted to nearest neighbour connectivity on a line, and the
addition of permutation layers. As this disadvantages devices with a completely
connected coupling map\footnote{Note that some compilation strategies may
identify that the \SWAP{} gates in the permutation layer may be removed for
devices with all-to-all connectivity. We avoid this dependence on the
compilation strategy by fixing the connectivity in the uncompiled circuit to be
all-to-all.} \cite{Debnath2016}, a property which would typically be an
advantage, we choose not to make this restriction here. Notice, however, that
naively compiling \square{} onto an architecture with nearest neighbour
connectivity on a line would result in the circuits of
\cite{cross2018validating}. This similarity makes a comparison between
experiments involving these circuits relevant. As a result, compiling \square{}
to superconducting devices (where connectivity is low) will generally result in
a circuit similar to those used in the quantum volume benchmark, as many
\SWAP{} operations are required regardless. 

In summary, by allowing for the most general gate set and connectivity,
\square{} provide a means to thoroughly interrogate all layers of the \stack{}.
This class of circuits is inspired by quantum volume circuits
\cite{cross2018validating} but are somewhat generalised to avoid device bias.
Because of this generality, and the linear increase in depth with number of
qubits, \square{} are complementary to the other classes introduced in this
work.


\subsection{\Deep{}: Pauli Gadgets}
\label{sec:pauli gadgets}

Pauli gadgets \cite{cowtan2019phase} are quantum circuits implementing an
operation corresponding to exponentiating a Pauli tensor. Sequences of Pauli
gadgets acting on qubits form \emph{product formula} circuits, most commonly
used in Hamiltonian simulation \cite{Berry2007}. Many algorithms employing
these circuits require fault-tolerant devices, but they are also the basis of
trial state preparation circuits in many variational algorithms, which are the
most promising applications of noisy quantum computers. 

A notable example of this in quantum chemistry is the physically-motivated
Unitary Coupled Cluster family of trial states used in the variational quantum
eigensolver (VQE) \cite{peruzzo2014variational, barkoutsos2018quantum}.  As
near-term quantum computers hold promise as useful tools for studying quantum
chemistry, we propose that the quality of an implementation of these gadgets is
a useful benchmark, and use them to define the \emph{deep} circuit class.

\defpar{}

These circuits are built as in \algref{alg:pauli gadgets}. They are constructed
from several layers of Pauli Gadgets, each acting on a random subset of $n$
qubits. In the worst case each Pauli Gadget will demand $4n+1$ gates: $2n$
Pauli gates, $2\brac{n-1}$ \CX{} gates, and one \RZ{} gate. In the construction
of \deep{} there is discrete randomness over the choice of Pauli string, $s$,
and continuous randomness over a rotation angle $\alpha$.

\begin{figure}
\begin{algorithm}[H]
    \caption{The pattern for building \deep{}.}
    \label{alg:pauli gadgets}
    \inout{Number of qubits, $n \in \mathbb{Z}$}
        {$\brac{4n - 1}\brac{3n + 1}$}
        {Circuit, $C$}
    \begin{algorithmic}[1]
        \Function{PhaseGadget}{$\alpha$, $\curlbrac{\tilde{q}_1 , ... ,
        \tilde{q}_p}$}
            \If{$p = 1$}
                \State Act $\RZ{} \brac{\alpha}$ on $\tilde{q}_1$
            \Else
                \State Act \CX{} between $\tilde{q}_1$ and $\tilde{q}_2$
                \State \Call{PhaseGadget}{$\alpha$, $\curlbrac{\tilde{q}_2 ,
                ...  , \tilde{q}_p}$}
                \State Act \CX{} between $\tilde{q}_1$ and $\tilde{q}_2$
            \EndIf
        \EndFunction
        \State
        \Function{Pauli}{$\curlbrac{\tilde{q}_1 , ... , \tilde{q}_p}$, $s$}
            \If{$s_1 = \X{}$}
                \State Act $\RX{} \brac{\frac{\pi}{2}}$ on $\tilde{q}_1$
            \ElsIf{$s_1 = \Y{}$}
                \State Act \H{} on the $\tilde{q}_p$
            \EndIf
            \State \Call{Pauli}{$\curlbrac{\tilde{q}_2, ... ,
            \tilde{q}_p}$, $s$}
        \EndFunction
        \State
        \Function{PauliGadget}{$\alpha$, qubits, $s$}
            \State \Call{Pauli}{qubits, $s$}
            \State \Call{PhaseGadget}{$\alpha$, qubits}
            \State Act the inverse of \Call{Pauli}{qubits, $s$}
        \EndFunction
        \State
        \State Initialise $n$ qubits, labelled $q_1 , ... , q_n$, in the state
        $\ket{0}$.
        \State
        \For{each layer t up to depth $3n + 1$}
            \State \Comment The contents of this for loop constitutes a
                \emph{layer}. The choice of the number of layers used here is
                discussed in \appref{app:far from uniform deep}.
            \State
            \State Select a random string $s \in
            \curlbrac{\I{},\X{},\Y{},\Z{}}^n$
            \State Generate random angle $\alpha \in \sqrbrac{0 , 2 \pi}$
            \State \Call{PauliGadget}{$\alpha$, $\curlbrac{q_i : s_i \neq
            \I{}}$, $s$}
            \State
        \EndFor
        \State
        \State Measure all qubits in the computational basis
    \end{algorithmic}
\end{algorithm}
\end{figure}

By establishing the exponential distribution of the output probabilities from
\deep{}, as we do in \appref{app:far from uniform deep}, we allow ourselves the
capacity to use \HOG{} and \XE{} as introduced in \secref{sec:metrics}. This
constitutes a novel extension of those approaches to application motivated
benchmarking, and the unique ability for us to benchmark application-motivated
circuits, using polynomially many samples from a device. This provides a novel
insight into the capacity of near-term hardware to implement quantum chemistry
circuits.

\motpar{}

Note that the circuits in this class differ from running the VQE end-to-end.
Focusing on the state preparation portion of a VQE circuit, we might deduce
performance of the \stack{} when running the VQE on a number of
molecules.\footnote{We do not explore the relationship between the performance
of a \stack{} when implementing \deep{} and when implementing VQE, but regard
it to be important future work.} The intuition being that if the state
preparation \emph{is} accurate, then the error in the expectation values of
measured observables will be due to errors in implementing those observables,
or the readout process itself.

In summary, we have introduced \deep{} to probe the performance of a \stack{}
when implementing common quantum chemistry applications. The \deep{} circuit
class is complimentary to the other classes featured in this work by having a
depth which grows quickly with the number of qubits.

\section{Quantum Computing Stack}
\label{sec:stack}

Each component of a \stack{} exerts an influence on overall performance, and
identifying the distinct impact of a particular component is often hard. To
disentangle these factors, we must clearly identify the components used during
benchmarking. Here we detail the components used to build the \stack{}s
explored in \secref{sec:results}. The diverse selection of components allows us
to investigate a variety of ways of building a \stack{}.


\subsection{Software Development Kits}

We use a combination of tools available via \pytket{} \cite{tketpaper,
pytketdocs} and \qiskit{} \cite{qiskit, qiskitdocs}. \pytket{} is a Python
module which provides an environment for constructing and implementing quantum
circuits, as well as for interfacing with \tket{}, a retargetable compiler for
near term quantum devices, featuring hardware-agnostic optimisation. \qiskit{}
is a open-source quantum computing software development framework for
programming, simulating, and interacting with quantum processors, which also
provides a compiler. Details of the versions of the software used are seen in
\tabref{tab:package version} of \appref{app:compiler passes}.

We use three parts of \qiskit{} in this work. First is the \emph{transpiler
architecture}, which enables users to define a custom compilation strategy by
executing a series of \emph{passes} on the input circuit, as discussed in
\secref{sec:stack compilers}. The second part of \qiskit{} we use is the
library of predefined passes. Finally, a \emph{provider} is used to access
hardware made available over the cloud by IBM Quantum. The provider enables
users to send circuits to hardware, retrieve results, and query the hardware
for its properties.\footnote{These properties include the graph connectivity,
single- and two-qubit error rates, and qubit $T_{1}$ and $T_{2}$ times.  Some
of the \emph{noise-aware} compilation strategies we use require knowledge of
these properties.}

Similarly, we use \pytket{} to generate and manipulate circuits in several
ways. Firstly we use the \tket{} compiler to construct compilation strategies
which optimise the input circuit for the target hardware, utilising predefined
passes available in \tket{}. Secondly we use \pytket{} to define abstract
circuits and to convert \tket{}'s native representation of the circuit into a
\qiskit{} \texttt{QuantumCircuit} object which is then dispatched to IBM
Quantum's systems for execution.

\subsection{Compilers}
\label{sec:stack compilers}

\emph{Compilers} provide tools to construct executable quantum circuits from
abstract circuit models. This is done by defining \emph{passes} which may
manipulate a representation of a quantum circuit, often by taking account of
limited connectivity architectures, or minimising quantities such as gate
depth, but need not perform any manipulation.\footnote{An example of this is a
pass which counts the gates in the circuit.} These passes are composed to form
\emph{compilation strategies} which should output executable quantum circuits.
Quantum compiling is an active area of research
\cite{Khatri2019quantumassisted, Harrow2002efficient,
Wallman2016noise, heckey2015compiler, murali2019noise, abhari2012scaffold,
McCaskey_2020}, and there are many pieces of software available for quantum
compiling. As noted above, in this work we use two: \tket{} and the compiler
available in \qiskit{}.

For the purposes of this work, the problem of quantum compilation is divided
into three tasks.
\begin{description}
	\item[Placement:] Determine onto which \emph{physical} qubits of a given
        device the \emph{virtual} qubits in the circuit's representation should
        be initially mapped.
    \item[Routing:] Modify a circuit to conform to the qubit layout of a
        specific architecture, for example, by inserting \SWAP{} gates to allow
        non-adjacent qubits to interact \cite{cowtan2019routing}. Circuits are
        often designed without the device's coupling map in mind, so this step
        is important \cite{arute2020quantum}.\footnote{While
        hardware-efficient anz\"atze can prove advantageous in task such as
        machine learning and quantum chemistry in the near term, in general
        they may be restrictive and so we consider the more general problem.}
    \item[Optimisation:] Optimise the circuit according to some cost metric,
        usually gate count or circuit depth.
\end{description}
Each of these tasks could consider such things as the trade-offs between the
connectivity of a particular subgraph of the device and the amount of crosstalk
present in that subgraph \cite{murali2020software}.

Both \pytket{} and \qiskit{} have multiple placement, optimisation, and routing
passes. We compare the performance of 5 compilation strategies built from these
passes. Two of them, \CQC{} and \IBM{}, compile the circuit \emph{without}
knowledge of the device's noise properties.  Another two, \CQCnoise{} and
\IBMnoise{}, do take noise properties into account.  As a base line, we
consider a simple compilation strategy from \pytket{} using only routing,
without optimisation or noise-awareness; we refer to this compilation strategy
as \NONE{}. We detail these schemes in \appref{app:compiler passes}. The main
difference between the noise-aware schemes is that \CQCnoise{} prioritises the
minimisation of gate errors during placement,\footnote{This is true of \pytket{}
0.3.0; as a result of this work, later versions of \pytket{} take into account
readout error.} whereas \IBMnoise{} prioritises readout and \CX{} errors
\cite{murali2019noise}.

\subsection{Devices}
\label{sec:backends}

We benchmark some of the devices made available over the cloud by IBM Quantum.
The devices we use are referred to by the unique names \york{}, \melb{},
\sing{} and \our{}. Each device has a set of \emph{native gates} which all
gates in a given circuit must be decomposed to. For all the devices considered
here, the native gates are: an identity operation, \I{}; 3 ``$u$-gates''
\cite{IBMgates}, as defined in \equref{eq:physical gates}; and a controlled-NOT
(\CX{}) gate.
\begin{align}
    \begin{split}
        \Uthree{} \brac{\theta, \phi, \lambda} & = \brac{
            \begin{array}{cc}
                \cos \brac{\frac{\theta}{2}} & -e^{i \lambda}
                \sin \brac{\frac{\theta}{2}} \\
                e^{i \phi} \sin \brac{\frac{\theta}{2}} & e^{i
                (\lambda + \phi)} \cos \brac{\frac{\theta}{2}}
            \end{array}
            }\\
        \Utwo{} \brac{\phi, \lambda} & = \Uthree{}
            \brac{\frac{\pi}{2},\phi,\lambda} \\
        \Uone{} \brac{\lambda} & = \Uthree{} \brac{0,0,\lambda}
    \end{split}
    \label{eq:physical gates}
\end{align}

Two of the device properties used by the noise-aware compilation strategies are
their \emph{connectivity} and \emph{calibration data}.  Information about the
connectivity of a device is contained in its \emph{coupling map} which, in the
cases of the devices studied here, are shown in \appref{app:coupling maps} and
summarised in \tabref{tab:graph properties}. A coupling map is a graph with a
vertex for each qubit, and edges between vertices when there exists coupling
between the corresponding qubits.

We expect the coupling map of a device will influence the device's performance
according to our benchmarks. Coupling maps with a high average \emph{degree}
(mean number of edges incident on each vertex) have more qubits directly
connected to one another, reducing the requirement for \SWAP{} operations that
would increase the circuit depth.  Coupling maps with low \emph{radius}
(minimax distance over all pairs of qubits) have qubits which are closer to one
another, again reducing the need for \SWAP{} gates. Coupling maps with a high
number of \emph{vertices} (i.e., qubits), enables higher-width circuits to be
run. Finally, coupling maps where the number of vertices equals the
\emph{minimum cycle length} (smallest number of edges per cycle over all
cycles) have the advantage that any routing operation has more paths available
to it, potentially allowing for some parallelisation.

\begin{figure*}
    \begin{subfigure}[b]{0.3\textwidth}
        \centering
        \begin{tikzpicture}
            \node (graph) at (0,0) {\includegraphics[height=\textwidth]
            {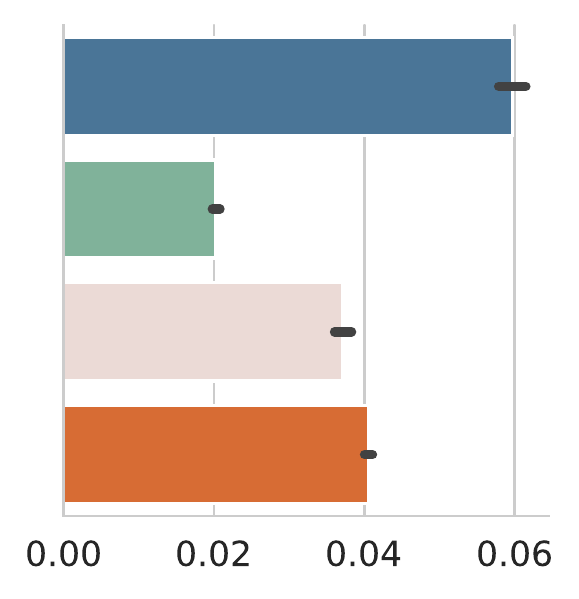}};

            \node[rotate = 90] at (graph.west) {Device};
        \end{tikzpicture}
        \caption{Readout Error.}
    \end{subfigure}
    \hfill
    \begin{subfigure}[b]{0.3\textwidth}
        \centering
        \begin{tikzpicture}
            \node (graph) at (0,0) {\includegraphics[height=\textwidth]
            {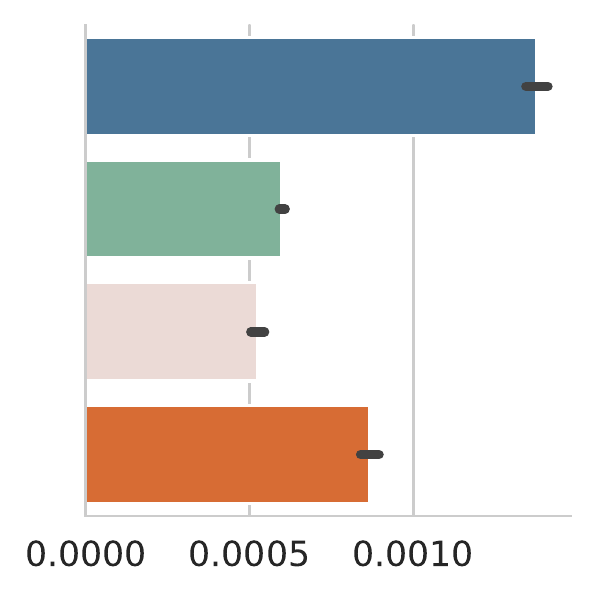}};

            \node at (graph.north) {Error Rate};
        \end{tikzpicture}
        \caption{Error per $\Utwo$ gate.}
    \end{subfigure}
    \hfill
    \begin{subfigure}[b]{0.3\textwidth}
        \centering
        \begin{tikzpicture}
            \node (graph) at (0,0) {\includegraphics[height=\textwidth]
            {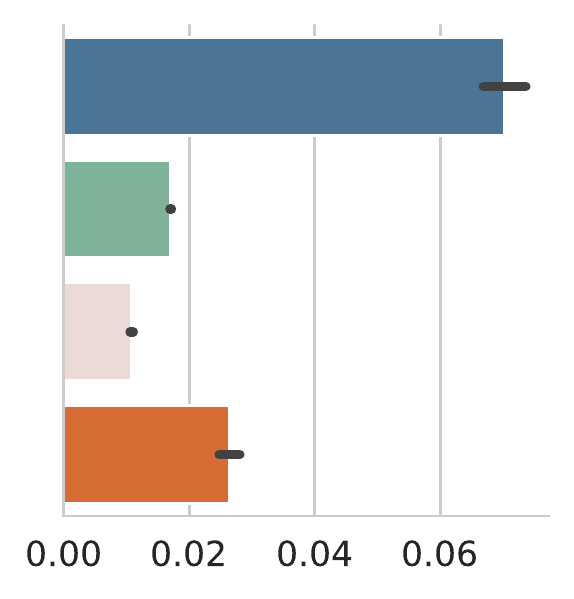}};
        \end{tikzpicture}
        \caption{Error per \CX{} gate.}
    \end{subfigure}
    \caption{\textbf{Average error rates across devices used in this work}.
    Bars show the mean error rates across the whole device, while error bars
    give the standard deviation. Devices shown here are: \york{}
    \crule{ibmqx2}, \our{} \crule{ibmq_ourense}, \sing{}
    \crule{ibmq_singapore}, \melb{} \crule{ibmq_16_melbourne}. Further details
    can be found in \appref{app:noise levels}}
    \label{fig:average noise}
\end{figure*}

Device calibration data includes information about single- and two-qubit error
rates, readout error, and qubit frequency, $T_{1}$, and $T_{2}$ times. The
noise-aware compilation strategies we investigate use the gate error rates and
readout error rates. Full details of noise levels can be found in \appref{app:noise
levels} with average values given in \figref{fig:average noise}. This
information is updated twice daily, with the data in \figref{fig:average noise}
averaged over the period 2020-01-29 to 2020-02-10 during which time our
experiments were conducted.

\begin{table*}
    \begin{tabularx}{\textwidth}{b s m s b}
        \toprule
        Device  & \#Vertices    & Average Degree    & Radius    & Minimum Cycle Length  \\
        \midrule
        \york{} & 5             & 2.4               & 1         & 3       \\
        \melb{} & 15            & $2 \frac{2}{3} $  & 4         & 4       \\
        \our{}  & 5             & 1.6               & 2         & N/A       \\
        \sing{} & 20            & 2.3               & 4         & 6     \\
        \bottomrule
    \end{tabularx}
    \caption{\textbf{Selected graph properties of the coupling maps of devices
    studied in this work}.  See \appref{app:coupling maps} for full details of
    the coupling maps of the devices explored here.}
    \label{tab:graph properties}
\end{table*}

The results of \secref{sec:results} depend heavily on the noise levels of
the device at the time at which the computation is implemented. This is doubly
true in the case of the noise-aware optimisation schemes as a circuit optimised
at one time may not perform as well over time as the noise levels of the devices
change. To reduce this effect we endeavoured to compile and run circuits within
as short a time interval as possible.

\section{Experimental Results}
\label{sec:results}

\newcommand\imgscale{0.6}

This section presents experimental results of running the circuit families
defined in \secref{sec:circuit classes} on various quantum computing systems.
We present a sub-sample of all our results in the figures which follow.  We
study these results in 3 contexts:
\begin{description}

    \item[\secref{sec:fullstack}:]\textbf{\fullstack{}} -- Incorporating and
        thoroughly investigating the compilation strategy helps develop an
        understanding of how circuit compilation influences the performance of
        the \stack{}.  For noise-aware compilation strategies, our results show
        that the assumptions made by the strategy about the importance of
        different kinds of noise impacts performance.

    \item[\secref{sec:appmotiv}:]\textbf{\appmotiv{}} -- By including three
        quite different circuit classes in our benchmark suite, we explore how
        a \stack{} may perform when implementing a wide array of applications.

    \item[\secref{sec:simins}:]\textbf{\simins{}} -- We explore how benchmarks
        assist in developing new noise models. By identifying when benchmark
        values for real implementations and those we expect from simulations
        using noise models differ, noise channels which should be added to the
        noise models to achieve greater agreement with real devices can be
        identified. This is of particular importance as noise-aware compilation
        strategies often utilise noise properties.

\end{description}
For each circuit class and fixed number of qubits, 200 circuits were generated
according to the algorithms of \secref{sec:circuit classes}.  Each circuit is
compiled by a given compilation strategy onto a particular device. The compiled
circuits were then run on the device, using 8192 repetitions (samples) from
each compiled circuit, which generates 8192 bitstrings. The compiled circuits
are also classically simulated using a noise model built from the device
calibration information at the time of the device run. See \emph{Data
Availability} for access to the full experimental data set.

The resulting bitstrings are then processed according to the figures of merit
given in \secref{sec:metrics}. The distribution of the figures of merit are
compared by their mean and distribution via a box-and-whisker plot.  In
particular boxes show quartiles of the dataset, whiskers extend to 1.5 times
the interquartile range past the low and high quartiles, and white circles give
the mean. Uncompiled circuits were also perfectly simulated without noise in
order to calculate the ideal heavy output probability. These points are
referred to as \emph{\noisefree{}} in the figures below.


\subsection{\fullstack{}}
\label{sec:fullstack}

\subsubsection{Impact of the Compilation Strategy}

We study the compilation strategy and the device on which the compiled circuit
is run. Using a fixed device and comparing multiple strategies allows us to
determine which strategy tends to perform well. Further, aggregating
performance over all compilation strategies provides an estimate of the
performance of a ``generic'' strategy.  Similarly, fixing the strategy, and
comparing its performance on multiple devices, shows how (possibly-erroneous)
assumptions made by the strategy about the devices impact performance.

\begin{figure*}
	\centering
	\begin{tikzpicture}
		\node (graph) at (0,0)
			 {\includegraphics[width=\imgscale\textwidth]{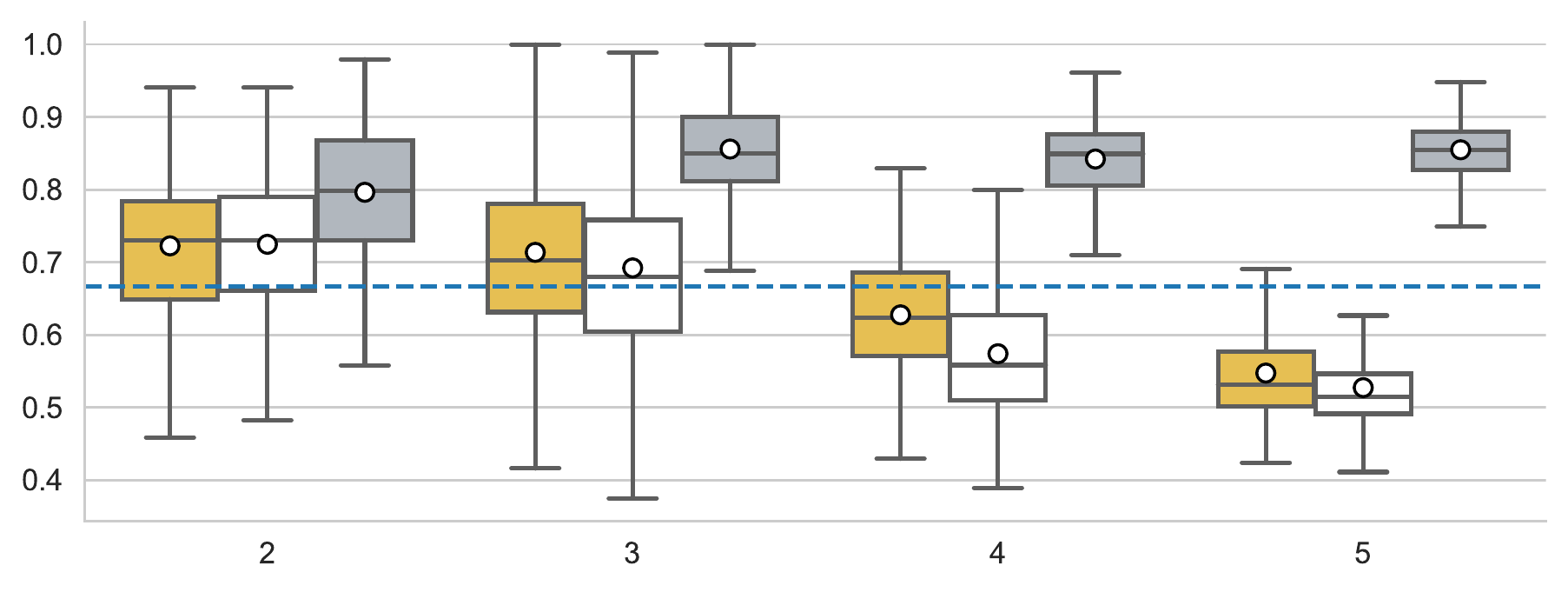}};

		\node at (graph.south) {Number of Qubits};
		\node[rotate=90] at (graph.west) {\titlecap{heavy outputs probability}};

		\node[align=left, anchor=west] at (graph.east) {
			\textbf{\underline{\uppercase{Strategy:}}}\\
			\crule{CQC-noise} \CQCnoise{} \\
			\crule{ideal} Noise Free \\
			\crule{all} All Strategies \\
		};
	\end{tikzpicture}
    \caption{\textbf{Comparison of fixed compilation strategy to average of all
    strategies using, the heavy outputs probability metric.} Here we have run
    \square{} using the real \melb{} device. Values above $2/3$ (dotted blue
    line), and closer to Noise Free, indicate better performance.}
    \label{fig:compiler_CQC-noise_vs_average_HO_square}
\end{figure*}

\figref{fig:compiler_CQC-noise_vs_average_HO_square} displays experimental
results when implementing \square{} on \melb{}, using heavy output generation
probability as the figure of merit. The \CQCnoise{} compilation strategy
performs somewhat better, on average, than a generic strategy. Because the
aggregated information (``All Strategies'' in
\figref{fig:compiler_CQC-noise_vs_average_HO_square}) includes aggregation over
\CQCnoise{}, these results indicate that other compilation strategies perform a
bit worse, since the performance of the aggregate is generally lower than that
of \CQCnoise{}.\footnote{Note that due to the fact we aggregate over all 5
compilation strategies, the distribution of heavy output probabilities amongst
the ``All Strategies'' category contains five times as many points as compared
to those for \CQCnoise{}.} This reveals the potential for compilation strategy
driven improvements in performance.

\begin{figure*}
	\centering
	\begin{tikzpicture}
		\node (graph) at (0,0)
			 {\includegraphics[width=\imgscale\textwidth]{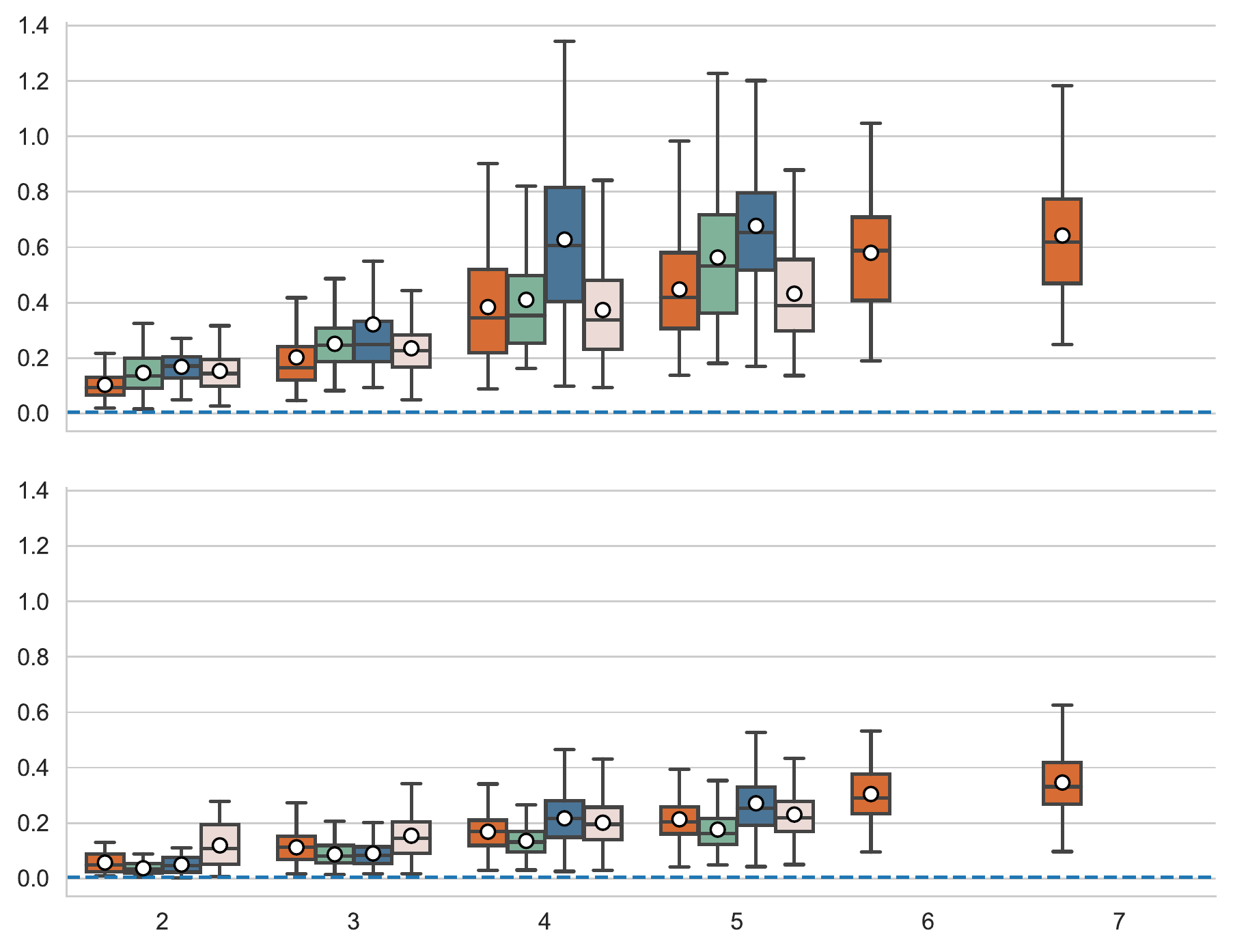}};

		\node at (graph.south) {Number of Qubits};
		\node at (0,0.1) {Classical Simulation Using Noise Model};
		\node at (graph.north) {Implementation on Real Backend};
		\node[rotate=90] at (graph.west) {\titlecap{\lone{}}};

		\node[align=left, anchor=west] at (graph.east) {
			\textbf{\underline{\uppercase{Device:}}}\\
			\crule{ibmq_singapore} \sing{} \\
			\crule{ibmqx2} \york{} \\
			\crule{ibmq_16_melbourne} \melb{} \\
			\crule{ibmq_ourense} \our{} \\
		};
	\end{tikzpicture}
    \caption{\textbf{Comparison of devices using the \lone{} metric, when
    running \shallow{} compiled using \CQCnoise{}}. Values close to $0$
    indicate better performance. Values below $1/192$ (dotted blue line) can be
    regarded as performing very well. Both simulations using \qiskit{} noise
    models, and implementations on real devices, are included.}
	\label{fig:arch_CQC-noise_l1_shallow}
\end{figure*}

\begin{figure*}
	\centering
	\begin{tikzpicture}
		\node (graph) at (0,0)
			 {\includegraphics[width=\imgscale\textwidth]{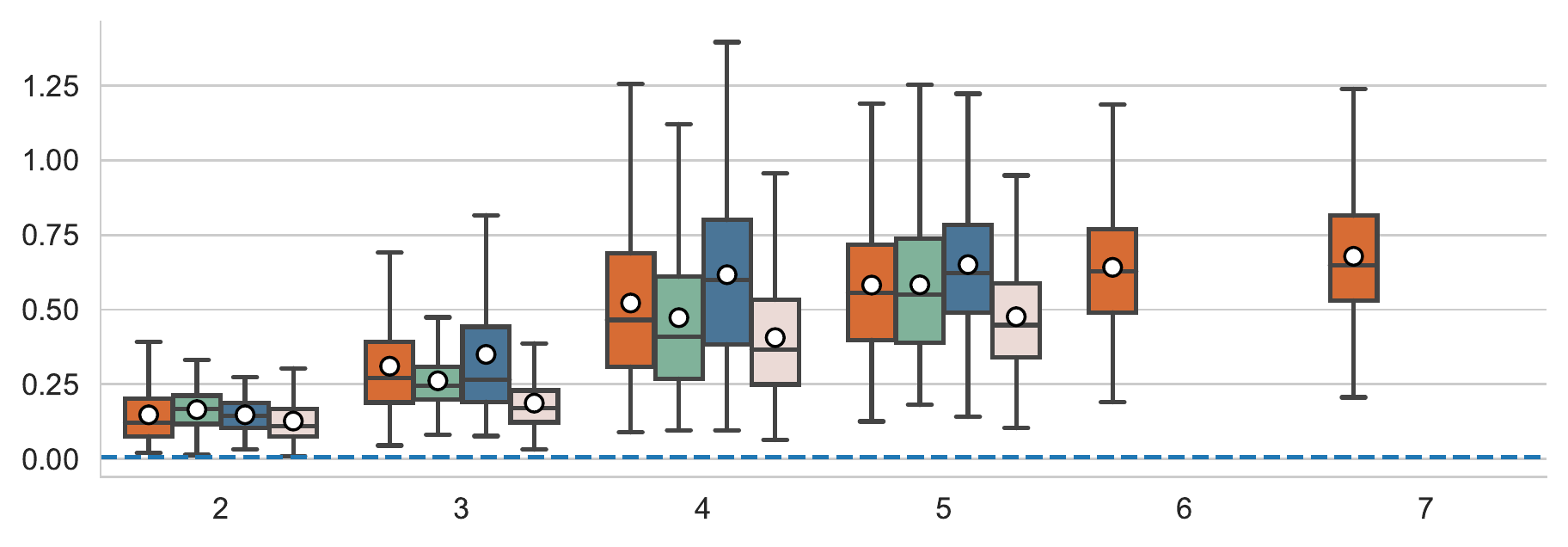}};

		\node at (graph.south) {Number of Qubits};
		\node[rotate=90] at (graph.west) {\titlecap{\lone{}}};

		\node[align=left, anchor=west] at (graph.east) {
			\textbf{\underline{\uppercase{Device:}}}\\
			\crule{ibmq_singapore} \sing{} \\
			\crule{ibmqx2} \york{} \\
			\crule{ibmq_16_melbourne} \melb{} \\
			\crule{ibmq_ourense} \our{} \\
		};
	\end{tikzpicture}
    \caption{\textbf{Comparison of real devices, using the \lone{} metric, when
    running \shallow{} compiled using all compilation strategies}. Here we
    compile onto each device using all compilation strategies, including all
    compiled circuits in this plot. Values close to $0$ indicate better
    performance. Values below $1/192$ (dotted blue line) can be regarded as
    performing very well.}
	\label{fig:arch_none_l1_shallow}
\end{figure*}

Aggregation over compilation strategies provides a way of identifying devices
which perform well, by ``washing out'' the effect of the compilation strategy
on performance.  \figref{fig:arch_CQC-noise_l1_shallow} shows that by
considering performance with a fixed compilation strategy \sing{} would be
considered to perform similarly, if not slightly better than \our{}, as
measured by \lone{}. However, aggregating over all strategies,
(\figref{fig:arch_none_l1_shallow}) shows \our{} to perform better, suggesting
that \our{} might be a better device for a ``generic'' compilation strategy to
compile to.

\begin{figure*}
	\centering
	\begin{tikzpicture}
		\node (graph) at (0,0)
			 {\includegraphics[width=\imgscale\textwidth]{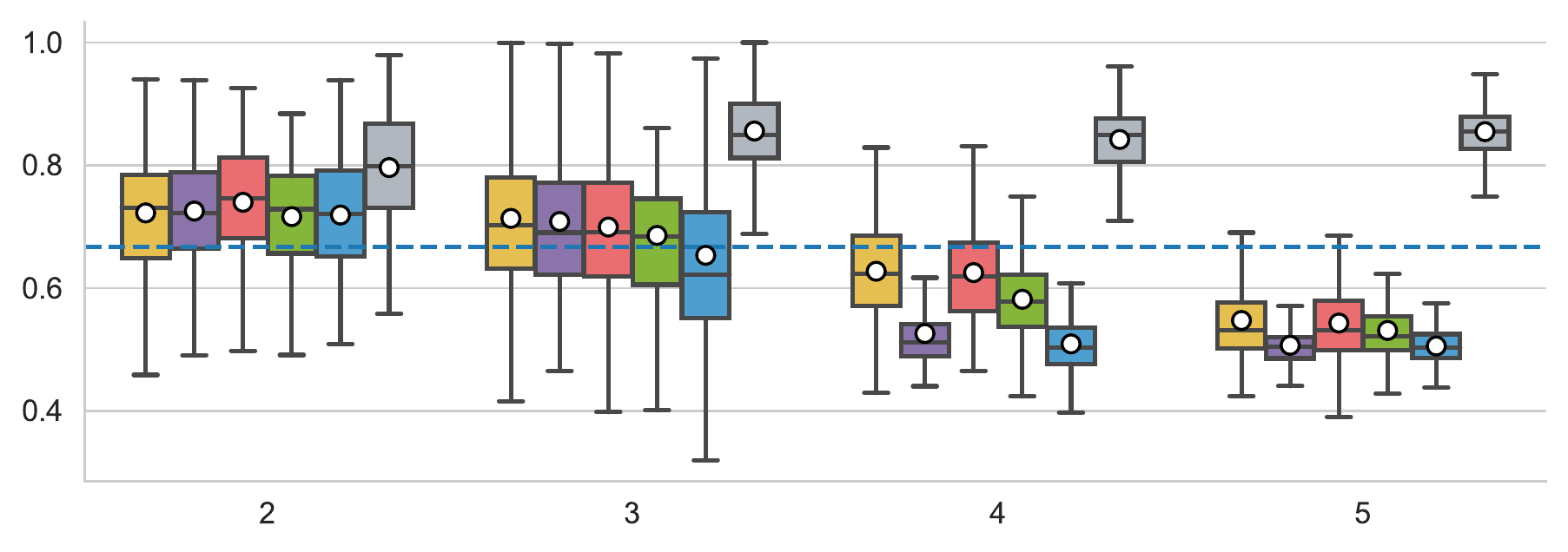}};

		\node at (graph.south) {Number of Qubits};
		\node[rotate=90] at (graph.west) {\titlecap{heavy outputs probability}};

		\node[align=left, anchor=west] at (graph.east) {
			\textbf{\underline{\uppercase{Strategy:}}}\\
			\crule{CQC-noise} \CQCnoise{} \\
			\crule{IBM-noise} \IBMnoise{} \\
			\crule{CQC} \CQC{} \\
			\crule{NONE} \NONE{} \\
			\crule{IBM} \IBM{} \\
			\crule{ideal} Noise Free \\
		};
	\end{tikzpicture}
    \caption{\textbf{Comparison of compilation strategies, using the heavy
    outputs probability metric, when running \square{} on the real \melb{}
    device}. Values above $2/3$ (dotted blue line), and closer to Noise Free,
    indicate better performance.}
	\label{fig:compiler_ibmq_16_melbourne_HO_square}
\end{figure*}

An instance-by-instance comparison of different compilation strategies also
reveals their limitations and advantages. For example,
\figref{fig:compiler_ibmq_16_melbourne_HO_square} shows \CQCnoise{} works best
at reproducing the ideal distribution of heavy output probabilities of
\square{} on \melb{}. When compared to the strong performance of \NONE{}, this
suggests these results are due in part to the routing scheme.

\begin{figure*}
	\centering
	\begin{tikzpicture}
		\node (graph) at (0,0)
			 {\includegraphics[width=\imgscale\textwidth]{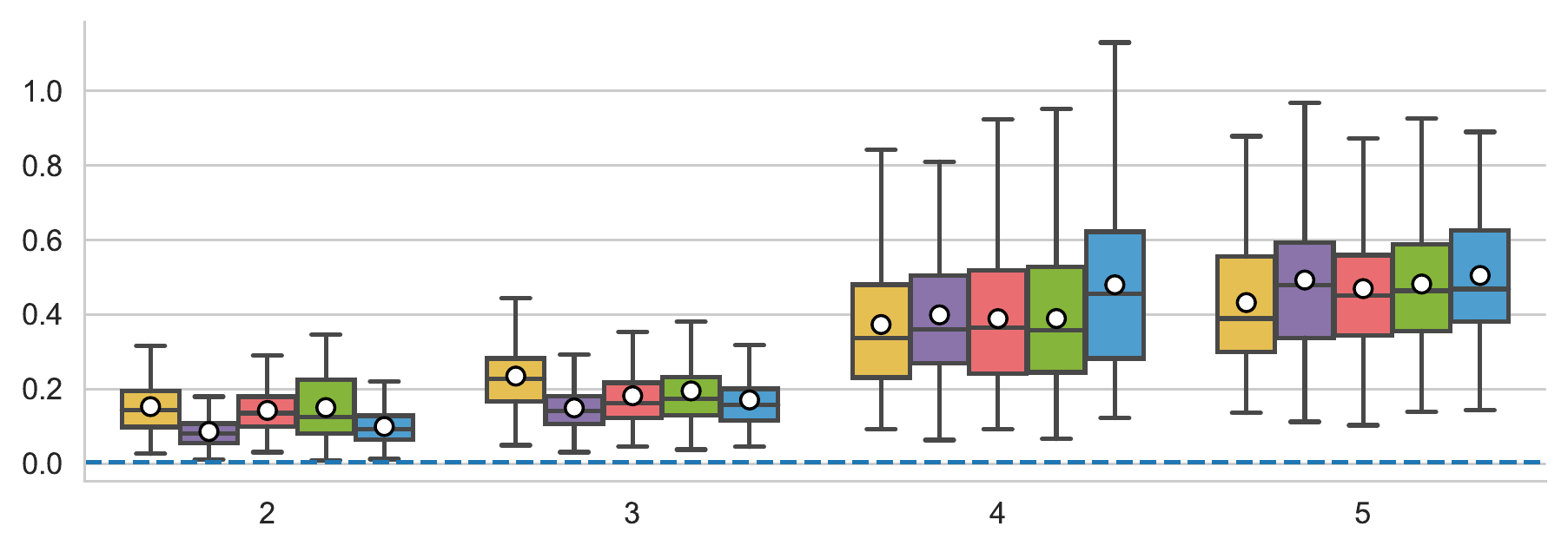}};

		\node at (graph.south) {Number of Qubits};
		\node[rotate=90] at (graph.west) {\titlecap{\lone{}}};

		\node[align=left, anchor=west] at (graph.east) {
			\textbf{\underline{\uppercase{Strategy:}}}\\
			\crule{CQC-noise} \CQCnoise{} \\
			\crule{IBM-noise} \IBMnoise{} \\
			\crule{CQC} \CQC{} \\
			\crule{NONE} \NONE{} \\
			\crule{IBM} \IBM{} \\
		};
	\end{tikzpicture}
    \caption{\textbf{Comparison of compilation strategies, using the \lone{}
    metric, when running \shallow{} on the real \our{} device}. Values close to
    $0$ indicate better performance. Values below $1/192$ (dotted blue line)
    can be regarded as performing very well.
    }
	\label{fig:compiler_ibmq_ourense_l1_shallow}
\end{figure*}

Similarly, \figref{fig:compiler_ibmq_ourense_l1_shallow} shows that \CQCnoise{}
is amongst the worst-performing compilation strategies for lower numbers of
qubits, while it is amongst the best-performing for higher numbers. This could
be a result of the way in which \CQCnoise{} prioritises noise in its routing
scheme, with gate errors taking precedence.\footnote{For larger numbers of
qubits and deeper circuits, gate errors becomes more impactful on the total
noise, and materialises as giving \CQCnoise{} an advantage for larger numbers
of qubits.}

\subsubsection{Noise Level, Connectivity Trade Off}

More highly-connected architectures typically allow for shallower
implementations of a given circuit as compared to less-connected ones. However,
the noise levels may be higher due to crosstalk \cite{PhysRevX.10.011022},
resulting in a trade-off between connectivity and the total amount of noise
incurred when running a computation.  Noise affects the accuracy of the
computation, so this trade-off has practical implications for the performance
of a device.  Reducing the connectivity between superconducting qubits has been
used to reduce noise levels \cite{PhysRevX.10.011022}. In superconducting
qubits, this can also be counteracted using tunable couplers \cite{Arute2019},
but this is not utilised in the devices studied here.\footnote{While we focus on
the connectivity of superconducting architectures here, more generally the
comparison between the limited connectivity of superconducting devices, and the
completely connected coupling maps of ion trap devices is of interest
\cite{Linke3305, PhysRevA.99.062323, Benedetti2019}.}

\begin{figure*}
	\centering
	\begin{tikzpicture}
		\node (graph) at (0,0)
			 {\includegraphics[width=\imgscale\textwidth]{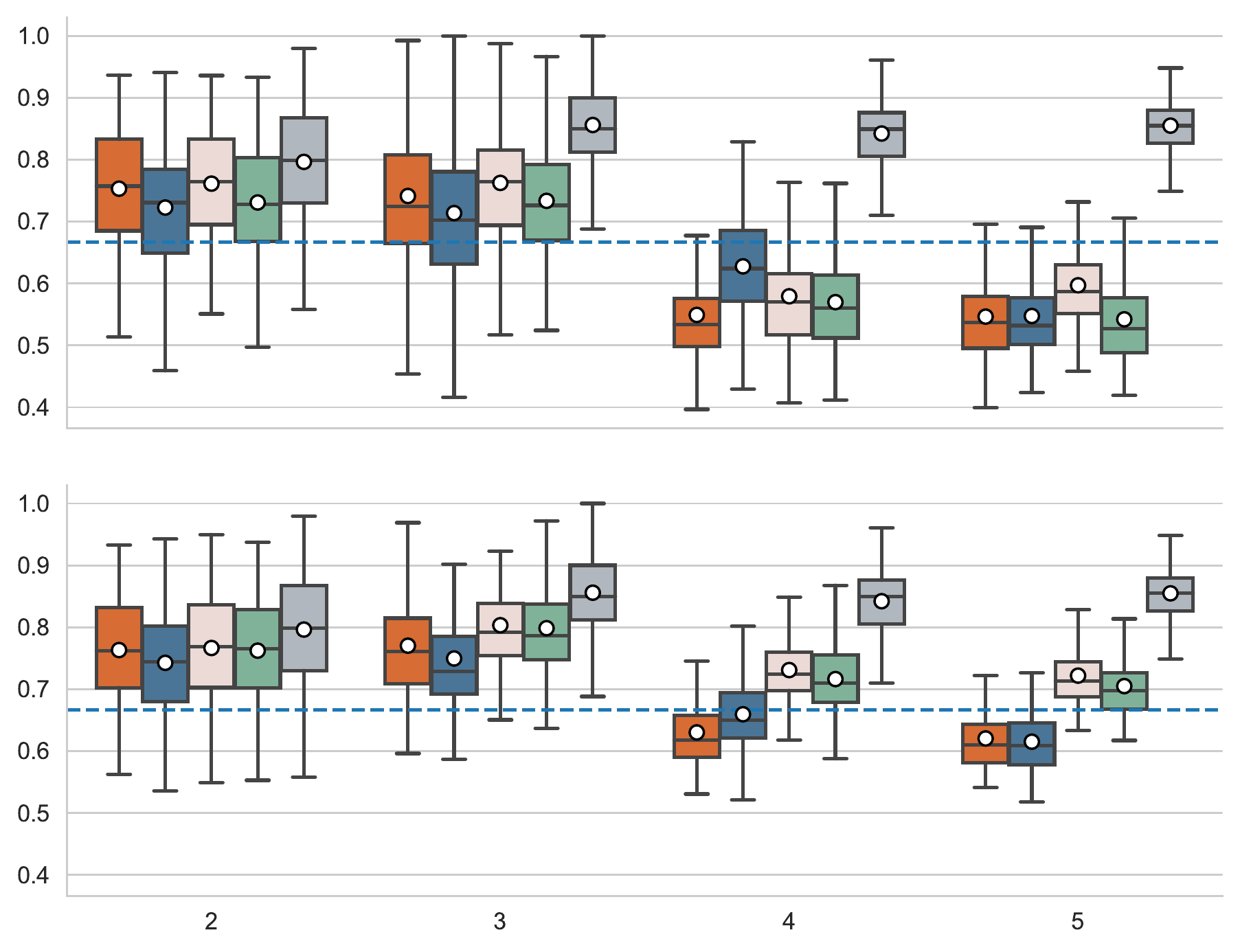}};

		\node at (graph.south) {Number of Qubits};
		\node at (0,0.1) {Classical Simulation Using Noise Model};
		\node at (graph.north) {Implementation on Real Backend};
		\node[rotate=90] at (graph.west) {\titlecap{heavy outputs probability}};

		\node[align=left, anchor=west] at (graph.east) {
			\textbf{\underline{\uppercase{Device:}}}\\
			\crule{ibmq_singapore} \sing{} \\
			\crule{ibmq_16_melbourne} \melb{} \\
			\crule{ibmq_ourense} \our{} \\
			\crule{ibmqx2} \york{} \\
			\crule{ideal} Noise Free \\
		};
	\end{tikzpicture}
    \caption{\textbf{Comparison of devices using the heavy outputs probability
    metric, when running \square{} compiled using \CQCnoise{}}. Values above
    $2/3$ (dotted blue line), and closer to Noise Free, indicate better
    performance. Both simulations using \qiskit{} noise models, and
    implementations on real devices, are included.}
	\label{fig:arch_CQC-noise_HO_square}
\end{figure*}

\begin{figure*}
	\centering
	\begin{tikzpicture}
		\node (graph) at (0,0)
			 {\includegraphics[width=\imgscale\textwidth]{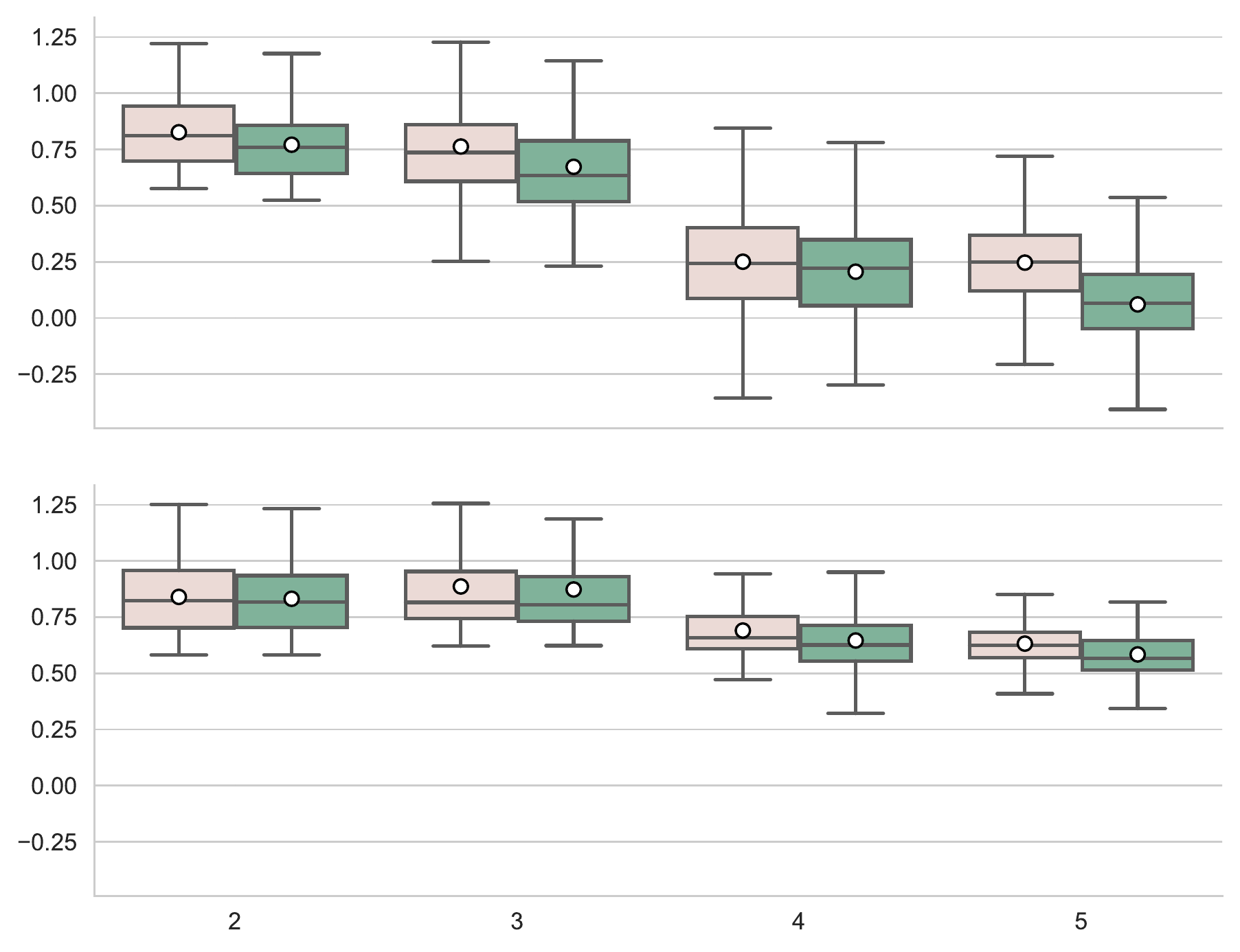}};

		\node at (graph.south) {Number of Qubits};
		\node at (0,0.1) {Classical Simulation Using Noise Model};
		\node at (graph.north) {Implementation on Real Backend};
		\node[rotate=90] at (graph.west) {\titlecap{cross entropy difference}};

		\node[align=left, anchor=west] at (graph.east) {
			\textbf{\underline{\uppercase{Device:}}}\\
			\crule{ibmq_ourense} \our{} \\
			\crule{ibmqx2} \york{} \\
		};
	\end{tikzpicture}
    \caption{\textbf{Comparison of devices using the cross entropy difference
    metric, when running \square{} compiled using \CQCnoise{}}. Values close to
    $1$ indicate better performance. Both simulations using \qiskit{} noise
    models, and implementations on real devices, are included.}
	\label{fig:arch_CQC-noise_XE_square}
\end{figure*}

\figref{fig:arch_CQC-noise_HO_square} shows that devices with lower noise
levels (\sing{} and \our{}) typically outperform devices with higher noise
levels (\york{} and \melb{}) despite the latter's higher connectivity.  An
interesting exception to this is for 4 qubits, where \melb{} performs best,
likely because of the 4-qubit cycles in its connectivity graph, as mentioned in
\tabref{tab:graph properties}. This reduces the \SWAP{} operations necessary
for implementing the circuit, reducing the overall circuit depth. This reveals
the increase in performance that can be expected when the connectivity of the
device and the problem instance are similar \cite{arute2020quantum}.  Similar
results hold for \XE{}, as shown in \figref{fig:arch_CQC-noise_XE_square}.

In general, we expect that circuits whose structure can naturally be mapped to
the connectivity of the device will generally perform well, whereas those which
cannot, will not. In general though, lower-noise devices will tend to perform
best.


\subsection{\appmotiv{}}
\label{sec:appmotiv}

\begin{figure*}
	\centering
	\begin{tikzpicture}
		\node (graph) at (0,0) 
			 {\includegraphics[width=\imgscale\textwidth]{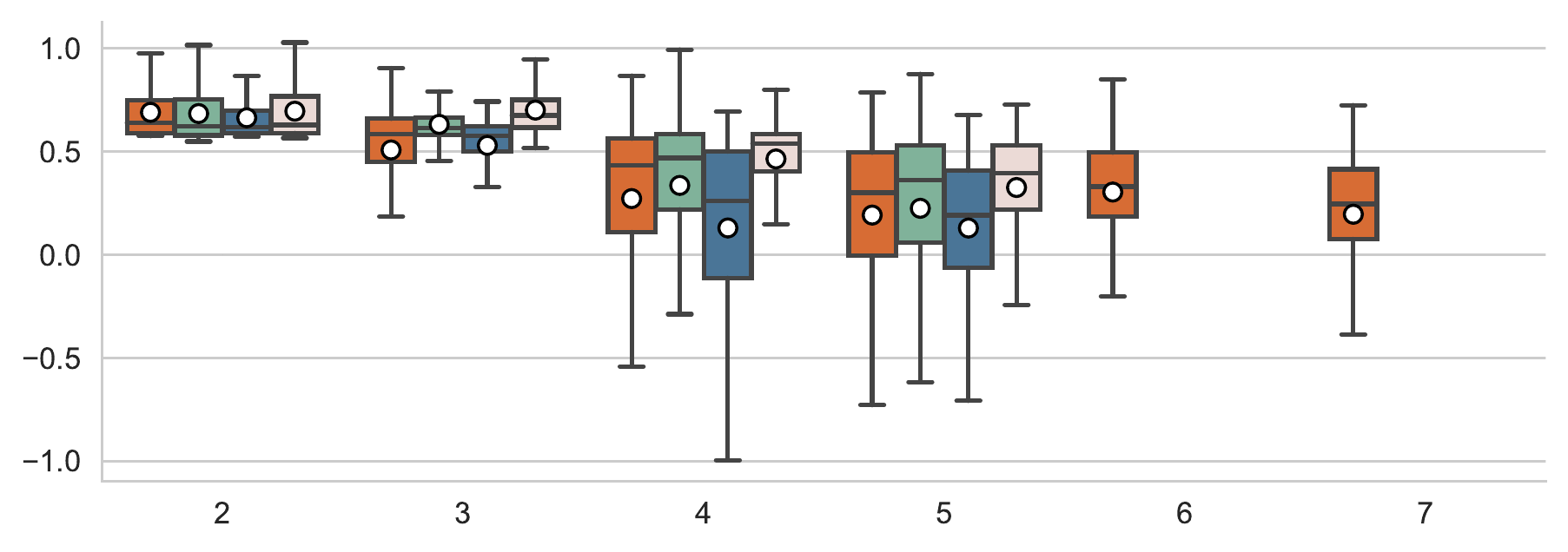}};

		\node at (graph.south) {Number of Qubits};
		\node[rotate=90] at (graph.west) {\titlecap{cross entropy difference}};

		\node[align=left, anchor=west] at (graph.east) {
			\textbf{\underline{\uppercase{Device:}}}\\
			\crule{ibmq_singapore} \sing{} \\
			\crule{ibmqx2} \york{} \\
			\crule{ibmq_16_melbourne} \melb{} \\
			\crule{ibmq_ourense} \our{} \\
		};
	\end{tikzpicture}
    \caption{\textbf{Comparison of real devices, using the cross entropy
    difference metric, when running \shallow{} compiled using \IBMnoise{}}.
    Values close to $1$ indicate better performance.}
    \label{fig:arch_IBM-noise_XE_shallow}
\end{figure*}

\begin{figure*}
	\centering
	\begin{tikzpicture}
		\node (graph) at (0,0)
			 {\includegraphics[width=\imgscale\textwidth]{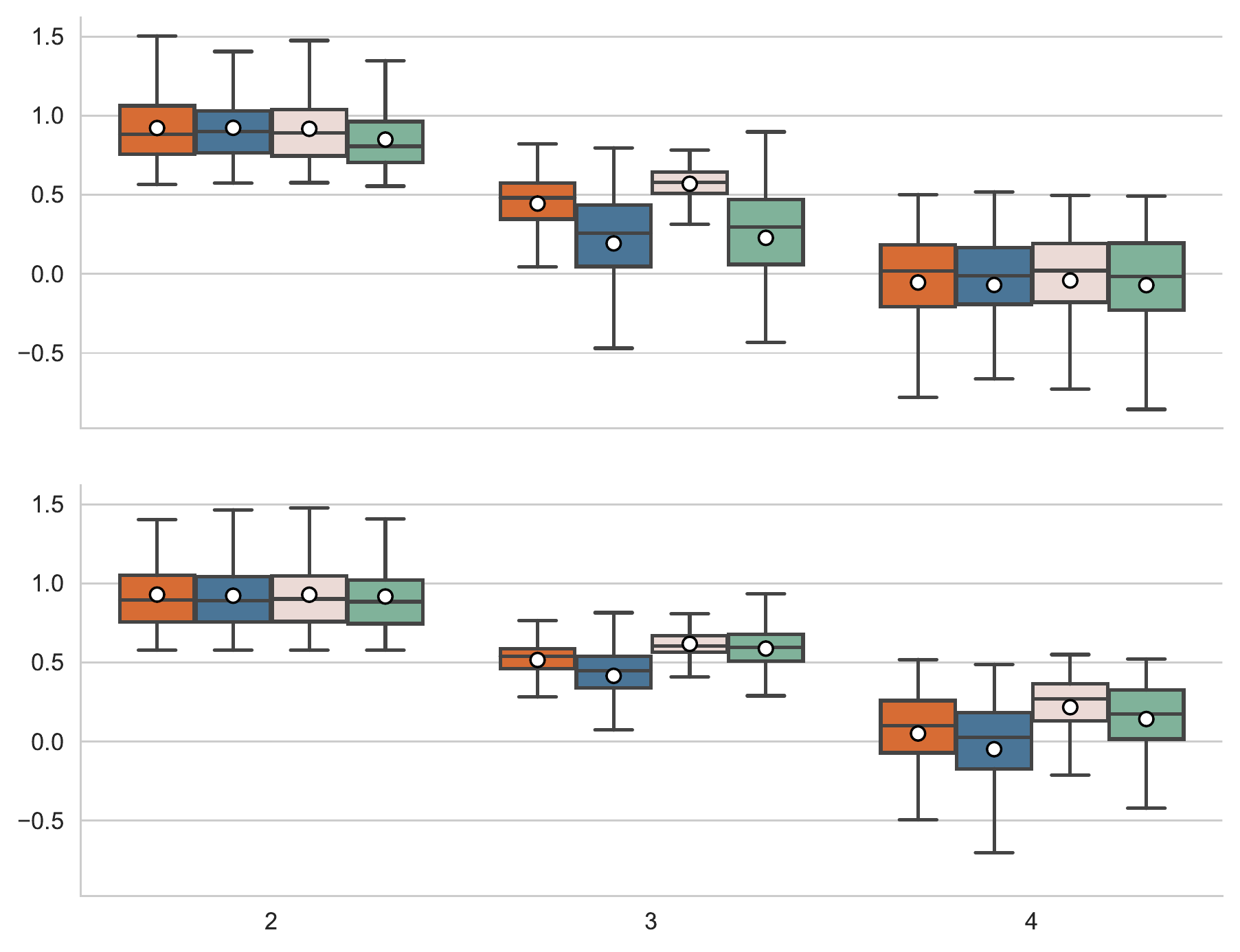}};

		\node at (graph.south) {Number of Qubits};
		\node at (0,0.1) {Classical Simulation Using Noise Model};
		\node at (graph.north) {Implementation on Real Backend};
		\node[rotate=90] at (graph.west) {\titlecap{cross entropy difference}};

		\node[align=left, anchor=west] at (graph.east) {
			\textbf{\underline{\uppercase{Device:}}}\\
			\crule{ibmq_singapore} \sing{} \\
			\crule{ibmq_16_melbourne} \melb{} \\
			\crule{ibmq_ourense} \our{} \\
			\crule{ibmqx2} \york{} \\
		};
	\end{tikzpicture}
    \caption{\textbf{Comparison of devices using the cross entropy difference
    metric, when running \deep{} compiled using \IBMnoise{}}. Values close to
    $1$ indicate better performance. Both simulations using \qiskit{} noise
    models, and implementations on real devices, are included.}
	\label{fig:arch_IBM-noise_XE_deep}
\end{figure*}

The same \stack{} will perform differently when running different applications,
as the structure of the circuits they require will generally be different.
Differences in performance are seen in the context of our application-motivated
benchmarks. For example, consider \figref{fig:arch_IBM-noise_XE_shallow}, which
shows performance when implementing sparsely connected circuits, and
\figref{fig:arch_IBM-noise_XE_deep}, which shows performance when implementing
chemistry-motivated circuits. In the case of
\figref{fig:arch_IBM-noise_XE_shallow}, the \york{} device outperforms \sing{},
while in the case of \figref{fig:arch_IBM-noise_XE_deep} the reverse is true.

\subsubsection{Quantum Chemistry}

\figref{fig:arch_IBM-noise_XE_deep} suggests \our{} is best for quantum
chemistry applications, because it performs well when running deep
circuits.\footnote{This comes with the caveat, as mentioned in \secref{sec:pauli
gadgets}, that the connection between the quality of an implementation of these
computational primitives, as measured by this benchmark, and accurate ground
state energy calculations in VQE has not been demonstrated experimentally.} In
particular \figref{fig:arch_IBM-noise_XE_deep} indicates that the average
circuit fidelity is highest for implementations on \our{}. 

In \figref{fig:arch_IBM-noise_XE_deep}, all devices converge to the minimum
value of cross-entropy difference at 4 qubits. To extend an investigation of
this sort to more qubits would require lower noise levels or chemistry
motivated circuits which generate exponentially distributed output
probabilities at lower depth.

\subsubsection{\titlecap{\shallow{}} as a Benchmark}

\begin{figure*}
	\centering
	\begin{tikzpicture}
		\node (graph) at (0,0)
			 {\includegraphics[width=\imgscale\textwidth]{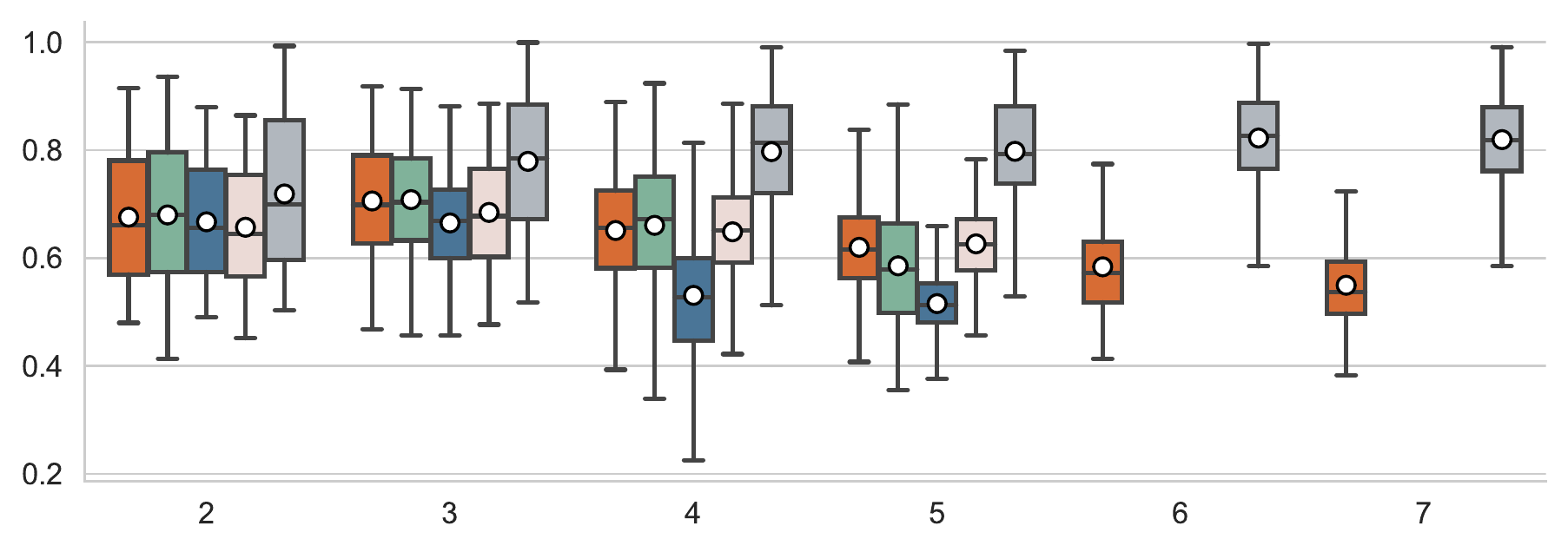}};

		\node at (graph.south) {Number of Qubits};
		\node[rotate=90] at (graph.west) {\titlecap{heavy outputs probability}};

		\node[align=left, anchor=west] at (graph.east) {
			\textbf{\underline{\uppercase{Device:}}}\\
			\crule{ibmq_singapore} \sing{} \\
			\crule{ibmqx2} \york{} \\
			\crule{ibmq_16_melbourne} \melb{} \\
			\crule{ibmq_ourense} \our{} \\
			\crule{ideal} Noise Free \\
		};
	\end{tikzpicture}
    \caption{\textbf{Comparison of real devices, using the heavy outputs
    probability metric, when running \shallow{} compiled using \CQCnoise{}}.
    Values close to Noise Free indicate better performance.}
    \label{fig:arch_CQC-noise_HO_shallow}
\end{figure*}

\figref{fig:arch_CQC-noise_HO_shallow} demonstrates that \shallow{} allow us to
benchmark the behaviour of a \stack{} for applications involving circuits with
many qubits but low circuit depth \cite{mills2017information, iqporig,
coyle2019born}. 

The results show \sing{} outperforms the comparably sized \melb{} and has
comparable performance to \our{} for smaller numbers of qubits. \sing{}
outperforms \our{} by having more qubits available. This superior performance
of \sing{} is in comparison to the results of
\figref{fig:arch_CQC-noise_HO_square}, where \our{} was shown to perform well.
This justifies our suggestion that \shallow{} should be included in
benchmarking suites. Doing so enables exploring higher-width circuits, and in
this setting devices that perform poorly when implementing \square{} or \deep{}
may perform well.


\subsection{\simins{}}
\label{sec:simins}

Noise in a non-fault-tolerant quantum computer results in discrepancies between
results obtained from running on real hardware and those that would be obtained
from an ideal quantum computer.  Noise models are utilised to help identify why
these discrepancies occur \cite{Vankov_2019}. However, a perfect model of the
noise -- which could reproduce the results of real hardware (up to statistical
error) -- could require many parameters to completely specify it. Therefore,
most noise models consider only a small handful of physically- motivated noise
sources.  Consequently discrepancies between the results of noisy simulation
and running experiments on real hardware always remain.

Historically, closing this gap required developing noise models of increasing
sophistication.  Doing so typically requires a great deal of physics expertise
to identify new noise channels. Further, new experiments would have to be
designed in order to estimate the parameters in the noise model.

Here, we show how application-motivated benchmarks can be helpful in
identifying whether new noise channels should be incorporated into a noise
model. By isolating the circuit types and coupling maps for which the
discrepancies are greatest, we gain intuition about the possible causes of the
mismatch.

For the devices explored here, the noise models are built using \qiskit{}. They
are derived from a device's properties, and include one- and two-qubit gate
errors\footnote{These are modelled to consist of a depolarising errors followed
by a thermal relaxation errors.} and single-qubit readout errors. We find these
noise models are inadequate to explain some of the discrepancies observed in
the data.

\subsubsection{Noise Does Not Just Flatten Distributions}

One discrepancy between experiments and noisy simulations is the spread of the
data. For example, \figref{fig:arch_CQC-noise_HO_square} shows that only in the
experimental case do the whiskers of the plot fall below the value $0.5$,
indicating the heavy outputs are less likely than they would be in the uniform
distribution. Some noise type, in particular one which shifts the probability
density, rather than uniformly flattening it, is not considered, or is under
appreciated, by the noise models used. Identifying that noise channel is left
to future work, though we speculate it may be related to a kind of thermal
relaxation error.

\subsubsection{Noise Models Under Represent Some Noise Channels}

The classical simulations in \figref{fig:arch_CQC-noise_HO_square} suggest
\york{} should perform similarly to \our{} in most cases. In fact, it quite
consistently performs worse. This is isolated in
\figref{fig:arch_CQC-noise_XE_square}, with the same phenomenon being observed
in \figref{fig:arch_CQC-noise_l1_shallow} and
\figref{fig:arch_IBM-noise_XE_deep}, showing the behavior is consistent across
all circuit types and figures of merit.

This difference between simulated and experimental results is pronounced in the
case of \figref{fig:arch_IBM-noise_XE_deep}, where \deep{} are used, suggesting
the noise models may be underestimating the error from time-dependent noises
such as depolarising and dephasing, or from two-qubit gates which are more
prevalent in \deep{}.

Another such example of a two-qubit noise channel, which is explicitly not
accounted for in our noise models, is crosstalk. The results in
\figref{fig:arch_CQC-noise_XE_square} are consistent with the expectation that
cross-talk should have the greatest impact on more highly connected devices
\cite{PhysRevX.10.011022}. As such crosstalk may be the origin of the
discrepancy. Of note is the fact this benchmark wasn't explicitly designed to
capture the effects of crosstalk, and yet those effects manifest themselves in
its results. We anticipate that including crosstalk-aware passes in compilation
strategies \cite{murali2020software} would reduce the discrepancy.

\section{Conclusion}
\label{sec:conclusion}


The performance of quantum computing devices is highly dependent on several
factors. Amongst them are the noise levels of the device, the software used to
construct and manipulate the circuits implemented, and the applications for
which the device is used. The impact of these factors on the performance of a
\stack{} are intertwined, making the task of predicting its holistic performance
from knowledge of the performance of each component impossible. In order to
understand and measure the performance of \stack{}s, benchmarks must take
this into consideration.


In this work we have addressed this problem by introducing a methodology for
performing application-motivated, holistic benchmarking of the full \stack{}. To
do so we provide a benchmark suite utilising differing circuit classes and
figures of merit to access a variety of properties of the device. This includes
the use of three circuit classes: \deep{} and \shallow{}, which are novel to
this paper; and \square{}, which resemble random circuits used in other
benchmarking experiments \cite{cross2018validating}. In addition we make use of
a diverse selection of figures of merit to measure the performance of the
\stack{}s considered, namely: \HOG{}, \XE{}, and the \lone{}. 

In particular, in the form of \deep{} we present an alternative to previous
approaches to application-motivated benchmarking. This is by considering
circuits inspired by one of the primitives utilised in VQE, namely Pauli gadgets
employed for state preparation, rather than VQE itself. Further, while we have
found that the performances of \stack{}s are indistinguishable when using
\square{} and \HOG{} for a large number of qubits, \shallow{} extend the number
of qubits for which detail can be observed, while also being consistent with
philosophy of volumetric benchmarking. 


We demonstrate this benchmark suite by employing it on \york{}, \melb{}, \our{},
and \sing{}. In doing so we justified our thesis that the accuracy of a
computation depends on several levels of the \stack{}, and that each layer
should not be considered in isolation. For example, identifying that the
increased connectivity of a device does not compensate for the increased noise,
as we do in \secref{sec:fullstack}, shows the impact of this layer of the
stack, and justifies investigating devices with a variety of coupling maps and
noise levels. By showing the differing performance between five compilation
strategies, we are able to identify, in \secref{sec:fullstack}, the dependence
of the best compilation strategy to use on the device and the dimension of the
circuit. This illustrates the dependence of the performance of the \stack{} on
the compilation layer, and the interdependence between the compilation strategy,
device and application on the overall performance of the \stack{}. In
particular, noise-aware compilation strategies often perform well, when the
noise model used by the strategy is accurate, as discussed in \secref{sec:simins}. 

In \secref{sec:appmotiv}, the wide selection of circuits within the proposed
benchmark suite reveals that the same device, evaluated according to a fixed
figure of merit, will perform differently when running different applications,
whose circuits are compiled by the same compilation strategy. Indeed the
comparative performance of (compilation strategy, device) pairs is shown to vary
between our circuit classes. This justifies our inclusion of circuit classes
which collectively cover a wide selection of applications in the benchmark suite
proposed here, and our full \stack{} approach.


We foresee the benchmarks conducted in this work providing a means to select the
best \stack{}, of those explored here, for a particular task, and vice versa. As
such we also anticipate that a variety of new \stack{}s could be benchmarked in
the way described in this work, empowering the user with knowledge about the
performance of current quantum technologies for particular tasks. 

These benchmarks may, in time, come to complement noise models and calibration
information as a means to disseminate information about a device's performance.
This parallels the use of the LINPACK benchmarks
\cite{petitet_whaley_dongarra_cleary} alongside FLOPS to compare diverse
classical computers. Recently, quantum volume, as defined in
\cite{cross2018validating}, has started to be adopted as one such metric
\cite{pino2020demonstration}, and we hope the benchmark suite developed here
will be incorporated similarly. Further, our benchmarks may facilitate an
understanding of how new, or hard-to-characterise, noise affects the practical
performance of quantum computers, as implied by the classical simulations of
\secref{sec:simins}.


The work presented here could be extended in several directions. The first is to
examine the impact of incorporating these benchmarks into a compilation
strategy.  While noise-aware compilation strategies currently use properties of
qubits to decide how to compile a circuit, it would be interesting to explore if
instead optimising for these benchmarks would change the compilation. The
trade off between the benefits of doing so against the increased compilation
time resulting from the time taken to perform the benchmarks should then also be
assessed. This information would help in the understanding of the interplay
between the amount of classical circuit optimisation performed and the amount by
which the performance of a quantum system can be increased.

Second, the philosophy of application-motivated benchmarking could be extended
to circuits which are more easily classically simulable. Because of their
reliance on classical simulation, the benchmarks introduced here may be used up
to, but not after, the point of demonstrating \supremacy{}. Hence new circuit
classes will need to be introduced which can be classically simulated in this
regime.  Alternatively, application-motivated benchmarks that are derived from
combining benchmarks of smaller devices \cite{Arute2019} could be developed.

Third, we envision a need to systematically study how properties of hardware,
such as noise levels or connectivity, influence a given device's performance.
In this work, we were limited to the particular devices made available by IBM
Quantum, which limits our ability to perform such a systematic inquiry. It is is
nevertheless vital to do so, as the results of \secref{sec:fullstack} show that
changing the hardware can dramatically influence performance. Indeed, this would
allow us to understand if the observations made in \secref{sec:fullstack} are
typical, and to explore the existence of other relationships. This could be
achieved by implementing this benchmark suite on more devices, or synthetic
devices with tunable coupling maps and noise information.

Finally, there is a need to study the correlation between the results of an
application-motivated benchmark and the performance of a \stack{} at running the
application which motivated it. This would show that benchmarking application
subroutines provides reliable predictors of performance when running the
application itself. While similar work has explored the correlation between the
classification accuracy and circuit properties of parametrised quantum circuits
\cite{hubregtsen2020evaluation}, comparing the performance of the benchmarks
defined here with their applications is a subject for future work. For example,
comparing the performance of a stack at implementing \deep{} and running the VQE
algorithm would show the extent to which \stack{}s that perform well at a
particular kind of state preparation circuit also perform well in estimating
properties of a wide range of molecules.

\subsubsection{Data Availability}

The data gathered during the experiments conducted for this work, as presented
in \secref{sec:results}, are available at \cite{mills_daniel_2020_3832121} -
\url{http://doi.org/10.5281/zenodo.3832121}.  QASM files representing the
circuits executed, the per sample bitstring outputs from device and simulator
executions, and device calibration data gathered throughout the course of the
experiments, are provided.

\subsubsection{Acknowledgements}

DM acknowledges support from the Engineering and Physical Sciences Research
Council (grant EP/L01503X/1). This work was made possible, in part, by systems
built by IBM as part of the IBM Quantum program, and made accessible via
membership in the IBM Q Network.

All authors would like to thank the anonymous reviewers for their
careful consideration of this work, and for their insightful comments and
suggestions.

IBM, IBM Q, and Qiskit are trademarks of International Business Machines
Corporation, registered in many jurisdictions worldwide. Other product or
service names may be trademarks or service marks of IBM or other companies.

\printbibliography

\appendix

\section{Exponential Distribution}
\label{app:far from uniform}

The \emph{exponential distribution}, with \emph{rate} $\lambda$, is a
probability distribution with the probability density function 
\begin{equation*}
	\mathrm{Pr} \brac{x} = \lambda e^{-\lambda x}.
\end{equation*}
This is the distribution of waiting times between events in a Poisson process.
We are concerned with showing that output probabilities of the circuit classes
considered here are exponentially distributed. Such a property is a signature
of quantum chaos, and that a class of circuits is approximately Haar random
\cite{boixo2018characterizing, Emerson2098, emerson2005convergence}. It also
allows for a simplified calculation of both the ideal value of the
cross-entropy discussed in \secref{sec:xed}, and the ideal heavy output
probability as discussed in \secref{sec:hog}, as discussed in those sections.
This in turn allows us to fully exploit \XE{} and \HOG{}. Here we will argue
numerically which of the circuits we introduce in \secref{sec:circuit classes}
generate output probabilities of this form,\footnote{This numerical approach to
demonstrating properties of distributions of output probabilities from
particular circuit classes parallels that taken in other work on benchmarking
\cite{aaronson2016complexity, boixo2018characterizing, bremejo2018architecture,
aaronson2011computational}.} and discuss the implications when they do not.  

We also demonstrate why the circuit depths used in \secref{sec:circuit classes}
are necessary to generate output probabilities of this form. To do this we
generate 100 circuits of each type and number of layers, where a layer is as
defined in the respective Algorithms of \secref{sec:circuit classes}. We then
calculate the ideal output probabilities using classical simulation and compare
this distribution of output probabilities to the exponential distribution. In
the case of \square{} and \deep{}, we notice a better approximation of the
exponential distribution by the distribution of output probabilities, measured
by the \lone{} between the two, as the number of layers increases. We can use
this to isolate the number of layers at which the difference approaches its
minimum. 

\subsection{\Square{}}
\label{app:far from uniform square}

The exponential form of the distribution of the output probabilities from
random circuits similar to \square{} has been established
\cite{aaronson2016complexity, boixo2018characterizing}. As the procedure we use
to generate \square{}, seen in \algref{alg:random circuits}, differs slightly
from that used for other similar random circuits \cite{aaronson2016complexity,
cross2018validating, boixo2018characterizing}, we explore the distribution of
its output probabilities here.

\begin{figure}
	\begin{subfigure}[b]{\columnwidth}
		\centering
		\begin{tikzpicture}
			\node (graph) at (0,0) 
				{\includegraphics[width=0.9\textwidth]{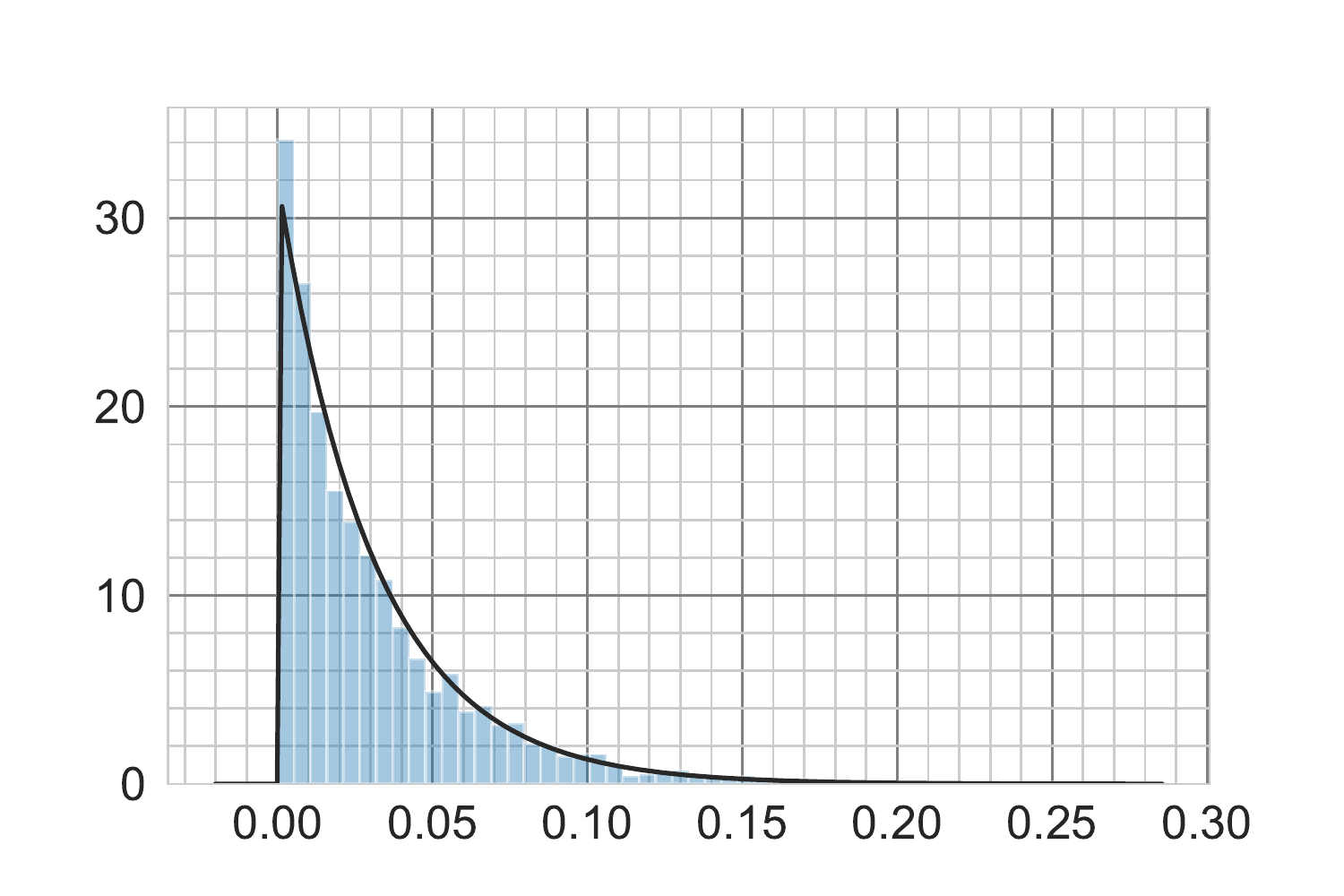}};
			\node at (graph.south) {$p_C$};
			\node[rotate=90] at (graph.west) {Probability Density};
		\end{tikzpicture}
		\caption{The distribution of output probabilities from a circuit
		$C$, where $C$ is a $5$ qubit circuit, from the \square{} class
		as defined in \algref{alg:random circuits}.} \label{fig:random exp
		dist fit}
	\end{subfigure}
	\begin{subfigure}[b]{\columnwidth}
		\centering
		\begin{tikzpicture}
			\node (graph) at (0,0) 
				{\includegraphics[width=0.9\textwidth]{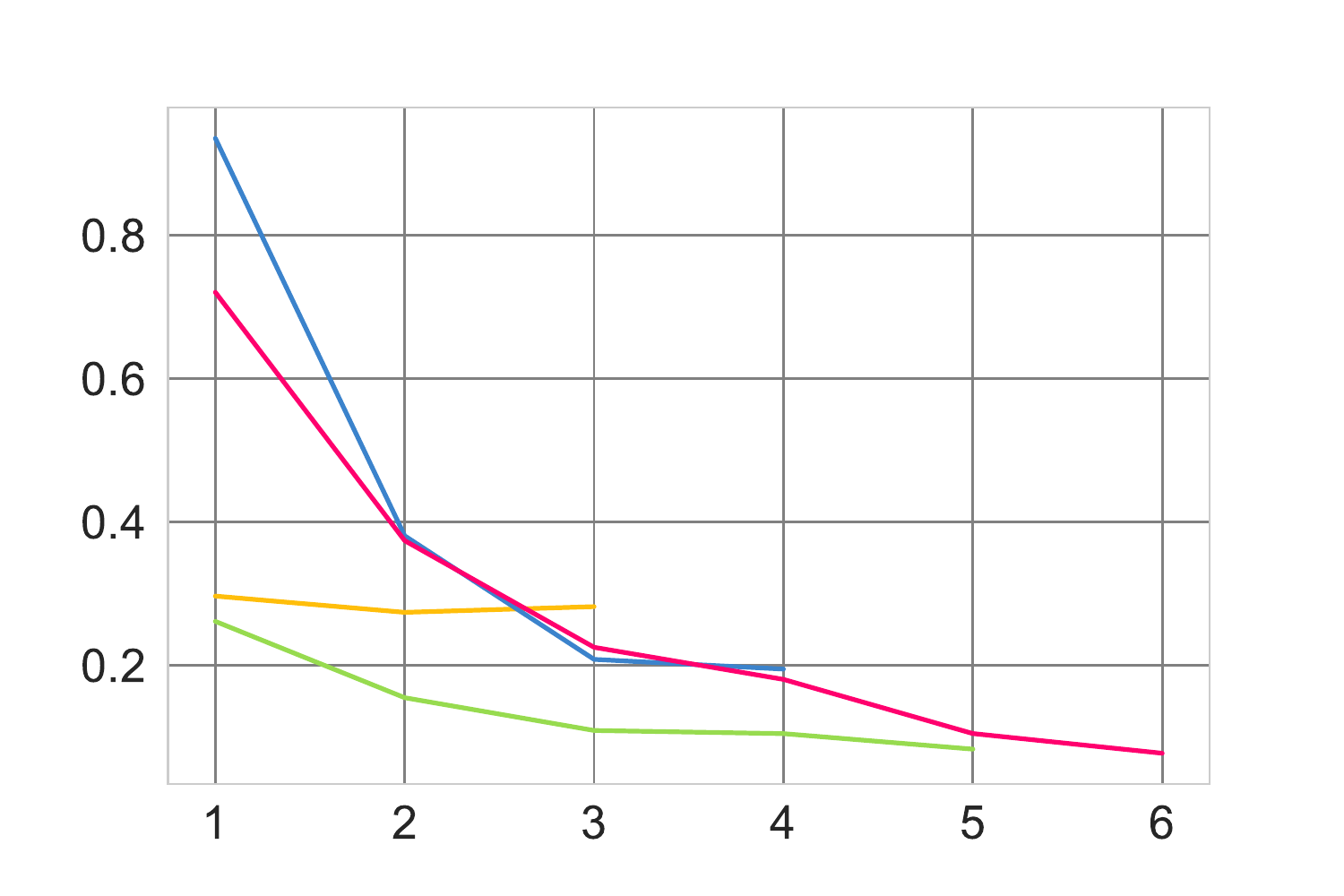}};
			\node at (graph.south) {Number of Layers};
			\node[rotate=90] at (graph.west) {\lone{}};
		\end{tikzpicture}
		\caption{The \lone{} between the distribution of output
		probabilities of \square{} and the exponential distribution
		$2^ne^{-2^nx}$, where $n$ is the number of qubits. A layer is
		defined as in \algref{alg:random circuits}. Colours correspond
		to numbers of qubits in the following way: 2 \crule{2}, 3
		\crule{3}, 4 \crule{4}, 5 \crule{5}.} 
		\label{fig:random exp dist convergence}
	\end{subfigure}
	\caption{\textbf{Exponential distribution fitting data for \square{}}.}
	\label{fig:random exp dist}
\end{figure}

The relevant results are seen in \figref{fig:random exp dist}. In particular,
it can be seen from \figref{fig:random exp dist convergence} that the minimum
value of \lone{} between the distribution of output probabilities and the
exponential distribution is approached at a number of layers equal to the
number of qubits, justifying our choice of layer numbers in \algref{alg:random
circuits}. It may be that asymptotically the number of layers required is
sub-linear \cite{boixo2018characterizing}, although for the circuit sizes used
here a linear growth in depth is appropriate.  \figref{fig:random exp dist fit}
illustrates the closeness of fit of the two distributions.

\subsection{\Deep{}}
\label{app:far from uniform deep}

Unlike with \square{}, there is no precedent for utilising \deep{} to generate
exponentially distributed output probabilities, as we do here. This allows us
to use \deep{} as a uniquely insightful benchmark of the performance of
\stack{}s, grounded both in the theoretical results of \secref{sec:metrics},
and in pertinent applications.

\begin{figure}
	\begin{subfigure}[b]{\columnwidth}
		\centering
		\begin{tikzpicture}
			\node (graph) at (0,0) 
				{\includegraphics[width=0.9\textwidth]{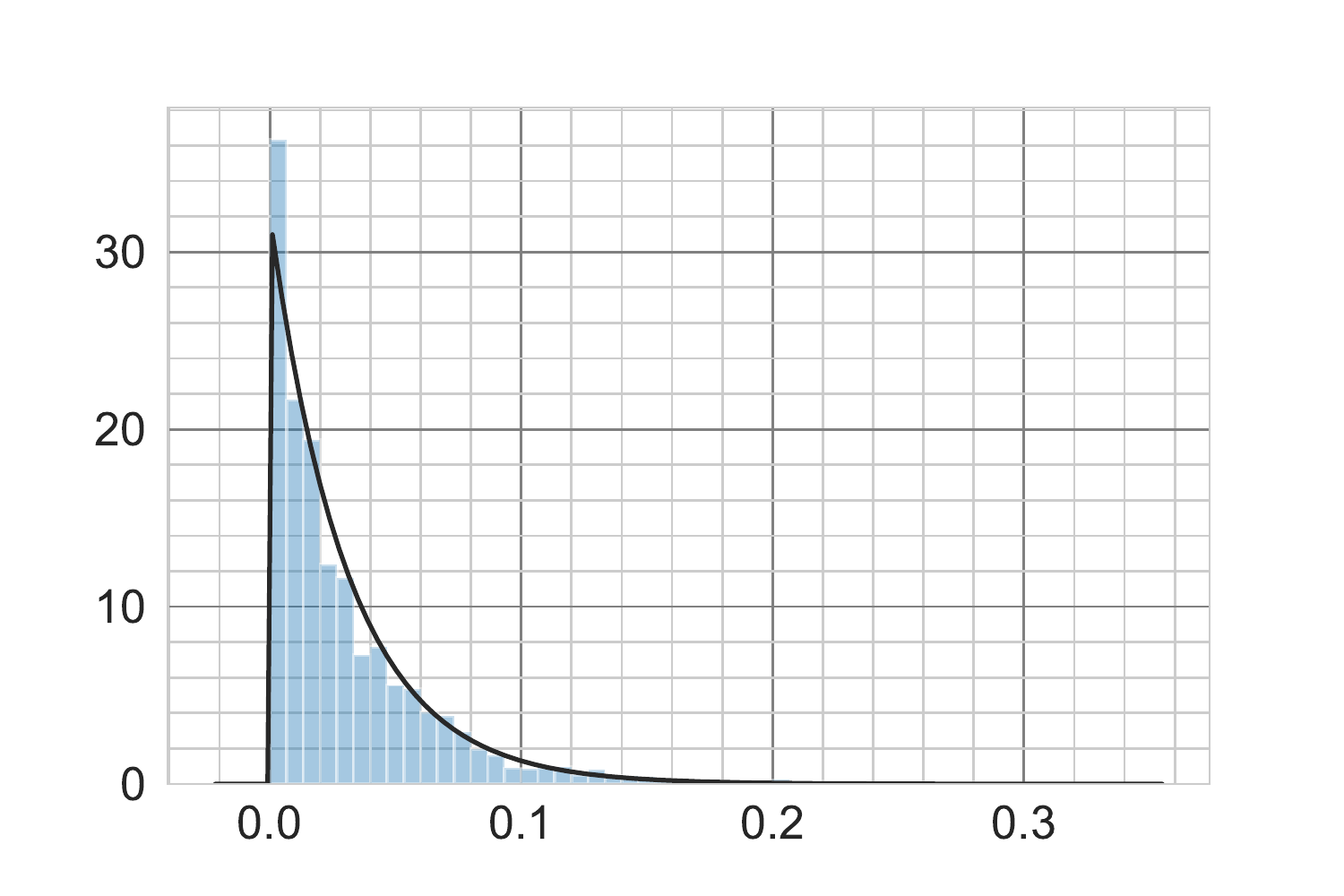}};
			\node at (graph.south) {$p_C$};
			\node[rotate=90] at (graph.west) {Probability Density};
		\end{tikzpicture}
		\caption{The distribution of output probabilities from a circuit
		$C$, where $C$ is a $5$ qubit circuit, from the \deep{} class as
		defined in \algref{alg:pauli gadgets}.} 
		\label{fig:pauli_gadg exp dist fit}
	\end{subfigure}
	\begin{subfigure}[b]{\columnwidth}
		\centering
		\begin{tikzpicture}
			\node (graph) at (0,0) 
				{\includegraphics[width=0.9\textwidth]{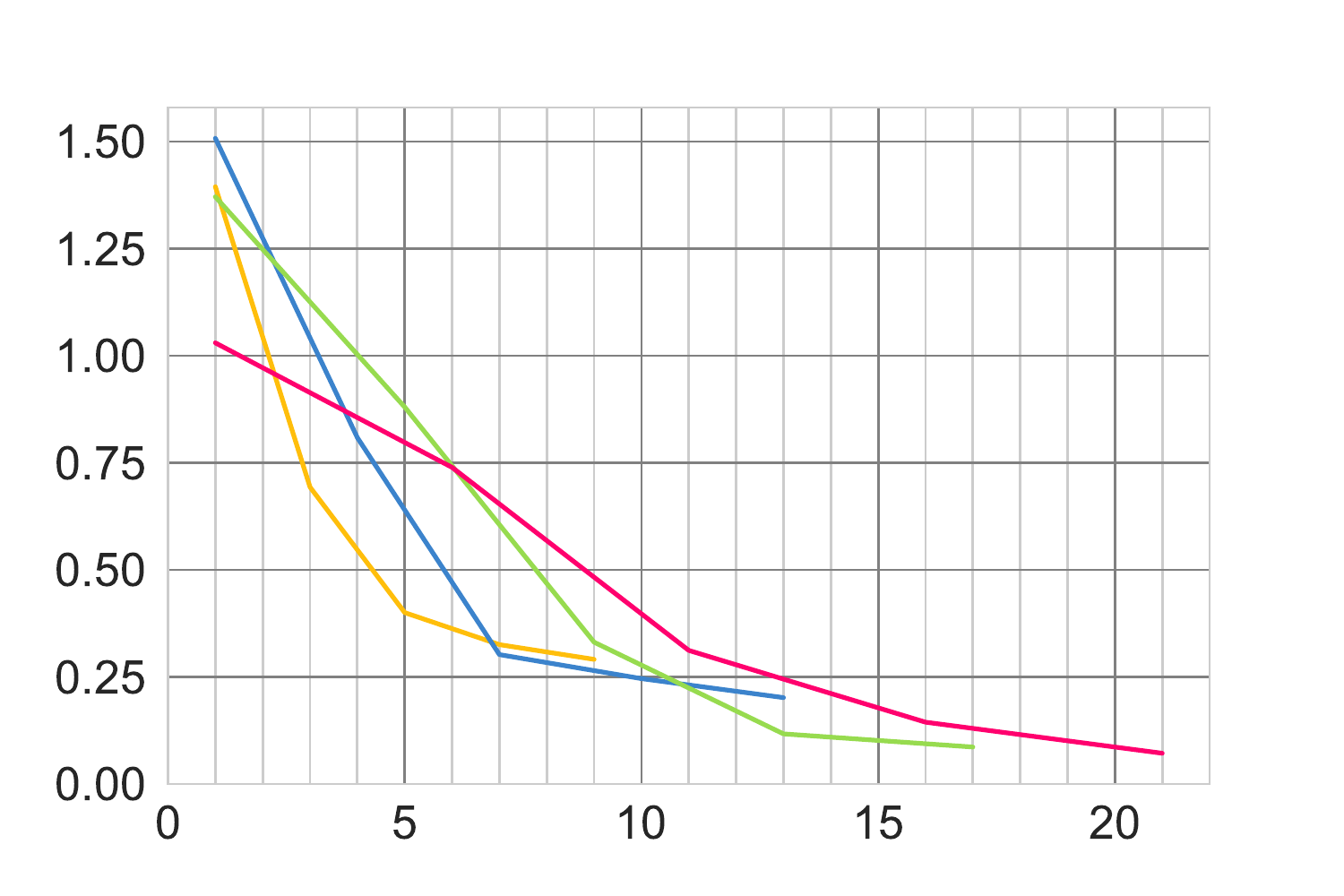}};
			\node at (graph.south) {Number of Layers};
			\node[rotate=90] at (graph.west) {\lone{}};
		\end{tikzpicture}
		\caption{The \lone{} between the distribution of output
		probabilities of \deep{} and the exponential distribution
		$2^ne^{-2^nx}$, where $n$ is the number of qubits. A layer is
		defined as in \algref{alg:pauli gadgets}. Colours correspond to
		numbers of qubits in the following way: 2 \crule{2}, 3
		\crule{3}, 4 \crule{4}, 5 \crule{5}.} 
		\label{fig:pauli_gadg exp dist convergence}
	\end{subfigure}
	\caption{\textbf{Exponential distribution fitting data for \deep{}}.}
	\label{fig:pauli_gadg exp dist}
\end{figure}

The relevant results are seen in \figref{fig:pauli_gadg exp dist}. In
particular, it can be seen from \figref{fig:pauli_gadg exp dist convergence}
that the minimum value of \lone{} between the distribution of output
probabilities and the exponential distribution is approached at a number of
layers equal to three times the number of qubits, plus one, justifying our
choice of layer numbers in \algref{alg:pauli gadgets}. \figref{fig:pauli_gadg
exp dist fit} illustrates the closeness of fit of the two distributions.

The depth required to achieve an exponential distribution of outcome
probabilities with \deep{} is greater than is the case for \square{}. Indeed,
random circuits were initially introduced as the shallowest circuits required
to generate such output probabilities \cite{boixo2018characterizing}. This
sacrifice in depth is made to achieve a benchmark which is uniquely application
motivated, as discussed in \secref{sec:circuit classes}.
	
\subsection{\Shallow{}}
\label{app:far from uniform shallow}

Unlike in the case of \square{} and \deep{}, the output probabilities of
\shallow{} are not exponentially distributed. This is unsurprising since random
circuits with this limited connectivity are thought to require at least depth
$\mathcal{O} \brac{\sqrt{n}}$ to create such a feature \cite{Brandao2016,
bouland2018quantum, bremejo2018architecture}. This has the unfortunate side
effect that the results of \secref{sec:xed} do not apply, and so the
simplifications to the calculations performed when conducting \XE{} cannot be
used. 

While it is also true that the predictions made about the ideal heavy output
probability, as discussed in \secref{sec:hog}, also do not apply, a study of
the heavy output probability is still of interest. In particular, while we
cannot connect the benchmark to the HOG problem of \probref{prob:hog}, we can
compare the probability of generating heavy outputs to the ideal probability of
producing heavy outputs, as calculated by classical simulation.

\section{Compilation Strategies} 
\label{app:compiler passes}

This section details the compilation strategies explored in each of our
experiments. For the circuit families and figures of merit investigated here,
the compilation strategies we used were designed and empirically confirmed to
perform well at the compilation tasks at hand. The version of each package used
are listed in \tabref{tab:package version}.  

\begin{table*}
	\begin{tabularx}{\textwidth}{@{}XX@{}} 
		\toprule 
		\textbf{Package}			            & \textbf{Version} \\ 
		\midrule 
		\qiskit{} \cite{qiskit, qiskitdocs}	    & 0.12.0 	\\ 
		\pytket{} \cite{tketpaper, pytketdocs}	& 0.3.0		\\ 
		\bottomrule 
	\end{tabularx} 
    \caption{\textbf{Packages used in this work, and their corresponding
    versions}.}
	\label{tab:package version} 
\end{table*}

\paragraph{\CQC{} and \CQCnoise{}} 

The \CQC{} and \CQCnoise{} compilation strategies are generated using
\algref{alg:cqc compiler pass}. \CQC{} is generated by passing $\texttt{False}$
as input to \algref{alg:cqc compiler pass}, and \CQCnoise{} by passing
$\texttt{True}$. 

Of particular interest are the following functions:
\begin{description}
	\item[\texttt{OptimiseCliffors}:] Simplifies Clifford gate sequences
		\cite{Fagan_2019}.
	\item[\texttt{KAKDecomposition}:] Identifies two-qubit sub-circuits with
		more than 3 \CX{}s and reduces them via the KAK/Cartan
		decomposition \cite{blaauboer2008analytical}. 
	\item[\texttt{route}:] Modifies the circuit to satisfy the architectural
		constraints \cite{cowtan2019routing}. This will introduce
		\SWAP{}	gates.
	\item[\texttt{noise\_aware\_placement}:] Selects initial qubit placement
		taking in to account reported device gate error rates
		\cite{tketpaper}.
	\item[\texttt{line\_placement}:] Attempts to place qubits next to those
		they interact with in the first few time slices. This does not
		take device error rates into account.
\end{description}

\begin{algorithm}[H]
	\caption{\pytket{} compilation strategies. The passes listed here are
	named as in the documentation for \pytket{} \cite{pytketdocs}, where
	additional detail on their actions can be found.}
	\label{alg:cqc compiler pass} 
	\vspace{7pt} 
	\hspace*{\algorithmicindent}
	\textbf{Input:}
	$\texttt{noise\_aware} \in \curlbrac{\texttt{True} , \texttt{False}}$ 

	\hrulefill
	\begin{algorithmic}[1] 
		\State \texttt{OptimiseCliffords}
		\State \texttt{KAKDecomposition} 
		\State
		\State \texttt{RebaseToRzRx} 
		\Comment{Convert to IBM gate set}
		\State \texttt{CommuteRzRxThroughCX} 
		\State 
		\If{\texttt{noise\_aware}}
			\State{\texttt{noise\_aware\_placement}}
		\Else
			\State \texttt{line\_placement} 
		\EndIf
		\State
		\State \texttt{route} 
		\State \texttt{decompose\_SWAP\_to\_CX} 
		\State \texttt{redirect\_CX\_gates} 
		\Comment{Orientate \CX{} to coupling map}
		\State 
		\State \texttt{OptimisePostRouting} 
		\Comment{Optimisation preserving placement and orientation}
	\end{algorithmic} 
\end{algorithm} 
	
\paragraph{\IBM{} and \IBMnoise{}} 

The \IBM{} and \IBMnoise{} compilation strategies, as defined in
\algref{alg:IBM compiler pass}, are heavily inspired by
\texttt{level\_3\_passmanager}, a preconfigured compilation strategy made
available in \qiskit{}. \IBM{} is generated by passing \texttt{noise\_aware} as
\texttt{False} in \algref{alg:cqc compiler pass}, and \IBMnoise{} by passing
\texttt{True}.

Where possible we passed \texttt{stochastic} as \texttt{True} in order to use
\texttt{StochasticSwap} instead of \texttt{BasicSwap} during the swap mapping
pass. In general, \texttt{StochasticSwap} generates circuits with lower depth;
however, for the versions listed in \tabref{tab:package version}, it proved
faulty for some circuit sizes and device coupling maps used in this work.
\texttt{StochasticSwap} may also result in repeated measurement of the same
qubit, which cannot be implement. Repeated compilation attempts may therefore
be necessary, and if this fails the circuit is not included in the plots of
\secref{sec:results}.

Of particular note are the following functions:
\begin{description}
	\item[\texttt{NoiseAdaptiveLayout}:] Selects initial qubit placement based
		on minimising readout error rates \cite{murali2019noise}.
	\item[\texttt{DenseLayout}:] Chooses placement by finding the most
		connected subset of qubits.  
	\item[\texttt{Unroller}:] Decomposes unitary operation to desired gate
		set.  
	\item[\texttt{StochasticSwap}:] Adds \SWAP{} gates to adhere to coupling
		map using a randomised algorithm.
	\item[\texttt{BasicSwap}:] Produces a circuit adhering to coupling map
		using a simple rule: \CX{} gates in the circuit which are not
		supported by the hardware are preceded with necessary \SWAP{}
		gates.
\end{description}

\begin{algorithm}[H]
	\caption{\qiskit{} compilation strategies. The passes listed here are
	named as in the documentation for \qiskit{} \cite{qiskitdocs}, where
	additional detail on their actions can be found.}
	\label{alg:IBM compiler pass}
	\vspace{7pt} 
	\hspace*{\algorithmicindent}
	\textbf{Input:} \\
	\hspace*{30pt} $\texttt{noise\_aware} \in \curlbrac{\texttt{True} , \texttt{False}}$ \\
	\hspace*{30pt} $\texttt{stochastic} \in \curlbrac{\texttt{True} , \texttt{False}}$ 

	\hrulefill
	\begin{algorithmic}[1] 
		\State \texttt{Unroller} 
		\State 
		\If{\texttt{noise\_aware}}
			\State \texttt{NoiseAdaptiveLayout}
		\Else 
			\State \texttt{DenseLayout}
		\EndIf
		\State \texttt{AncillaAllocation} 
		\Comment{Assign idle qubits as ancillas}
		\State
		\If{\texttt{stochastic}}
			\State \texttt{StochasticSwap}
		\Else
			\State \texttt{BasicSwap}
		\EndIf
		\State
		\State \texttt{Decompose(SwapGate)} 
		\Comment{Decompose \SWAP{} to \CX{}}
		\State \texttt{CXDirection} 
		\Comment{Orientate \CX{} to coupling map}
		\State
		\State
		\Comment{Gather 2 qubit blocks} 
		\State \texttt{Collect2qBlocks} 
		\State \texttt{ConsolidateBlocks}
		\State
		\State \texttt{Unroller} 
		\Comment{Unroll two-qubit blocks}
		\State \texttt{Optimize1qGates}
		\Comment{Combine chains of one-qubit gates}
		\State \texttt{CXDirection}
	\end{algorithmic} 
\end{algorithm} 

\paragraph{\NONE{}} 

In this case we perform, in the order as listed, the \pytket{} operations:
\texttt{route}, \texttt{decompose\_SWAP\_to\_CX}, and
\texttt{redirect\_CX\_gates}. We then account for the architecture gate set,
without any further optimisation.

\section{Device Data}
\label{app:device data}

Two device properties leveraged by our compilation strategies are the
\emph{coupling maps}, describing the connectivity of the qubits and in which
directions \CX{} gates can be performed, and the \emph{calibration
information}, describing the noise levels of the device. These properties, and
devices noise levels in particular, are considered valuable benchmarks of the
performance of the device in their own right.

These properties are collectively influential in noise-aware compiling, as
detailed in \appref{app:compiler passes}. There circuits are compiled to adhere
to the device's coupling map, while also aiming to minimise some function of
the calibration information. Because full \stack{} holistic benchmarking
encompasses the circuit compilation strategies, it provides a novel way of
using device information to benchmark an entire system, instead of simply the
physical qubits which comprise it.

\subsection{Device Coupling Maps}
\label{app:coupling maps}

\begin{figure*}
	\centering
	\begin{subfigure}[b]{0.24\textwidth}
		\centering
		\includegraphics[width=\textwidth]{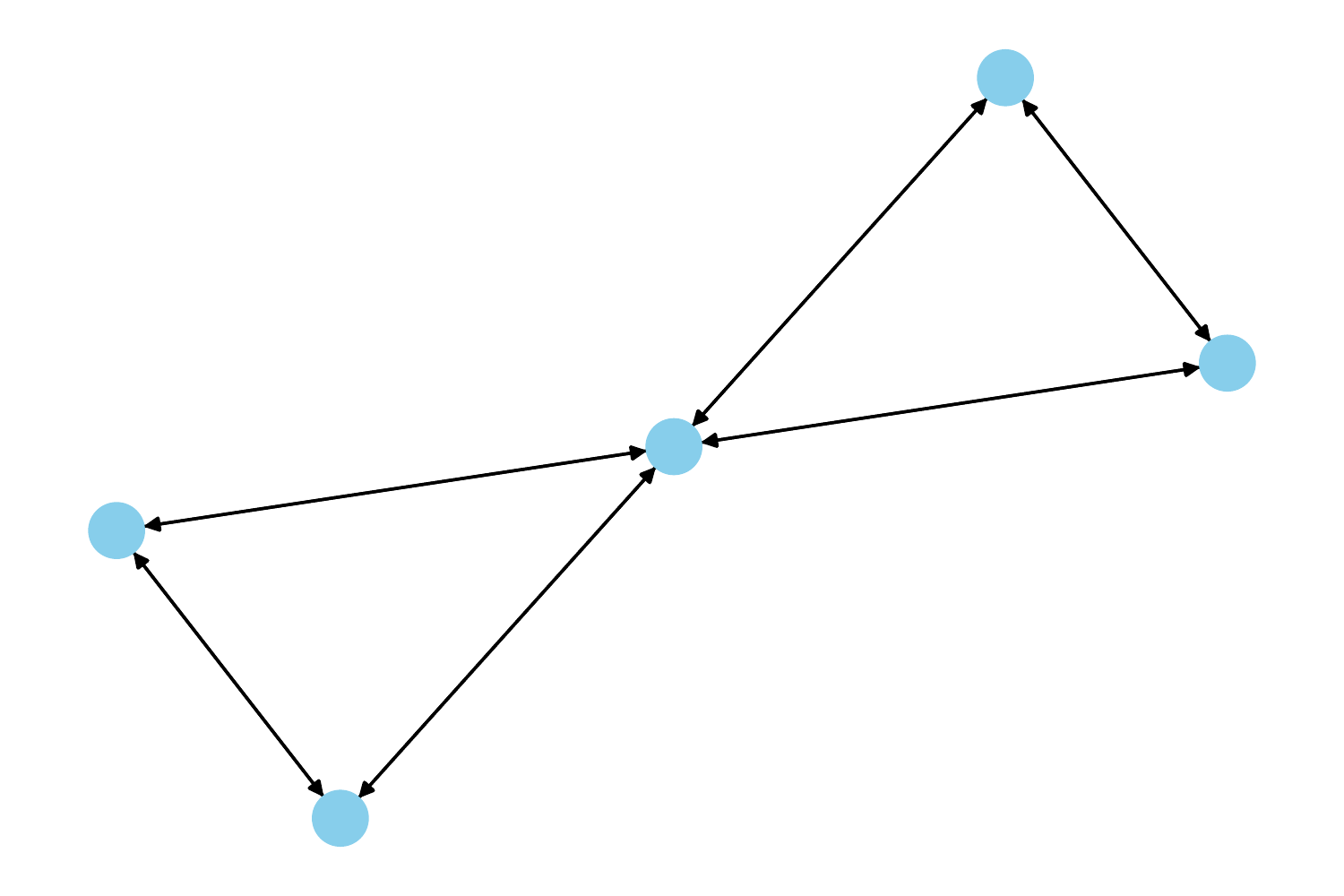}
		\caption{\york{}}
		\label{fig:ibmq_5_yorktown coupling map}
	\end{subfigure}
	\hfill
	\begin{subfigure}[b]{0.24\textwidth}
		\centering
		\includegraphics[width=\textwidth]{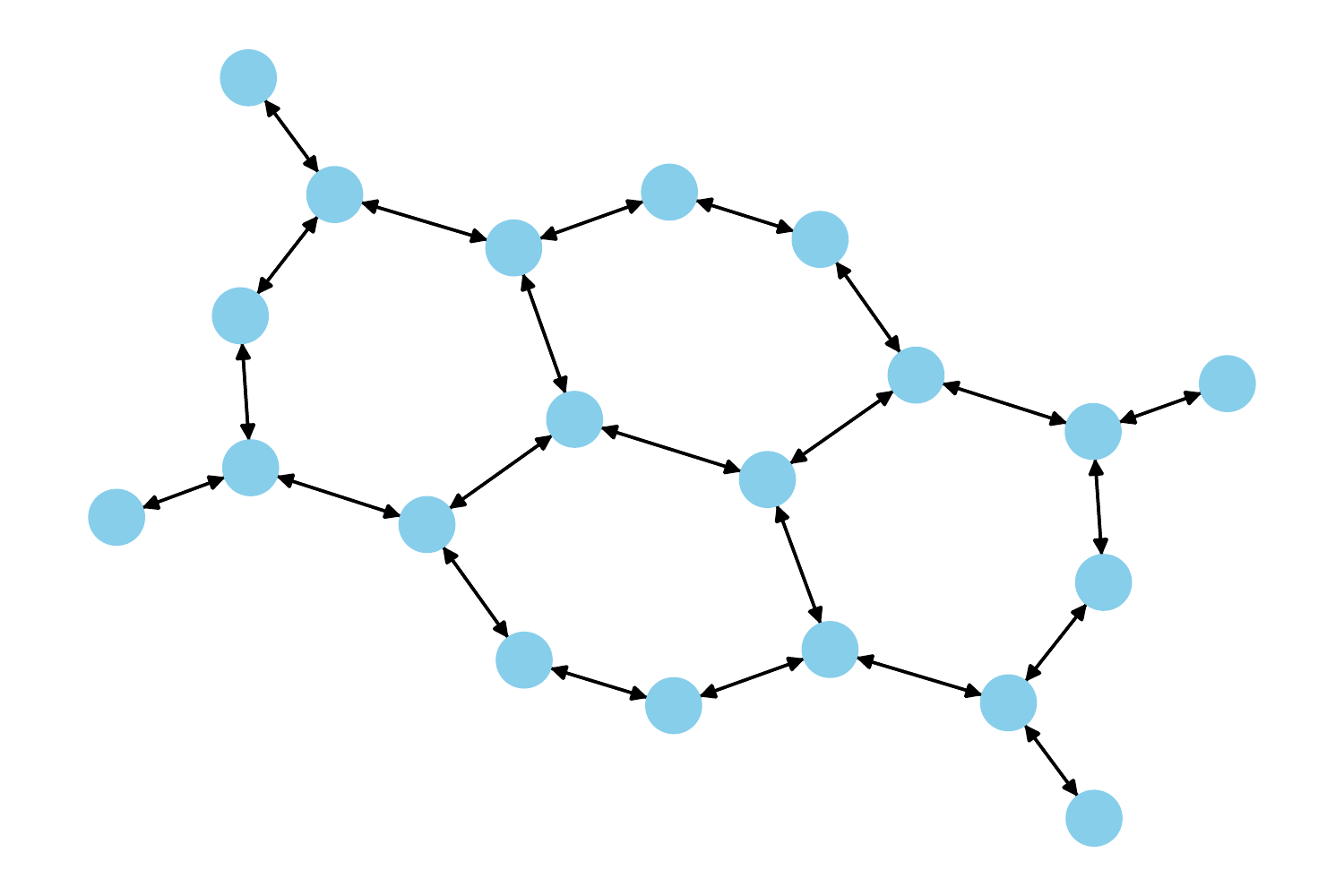}
		\caption{\sing{}}
		\label{fig:ibmq_singapore coupling map}
	\end{subfigure}
	\hfill
	\begin{subfigure}[b]{0.24\textwidth}
		\centering
		\includegraphics[width=\textwidth]{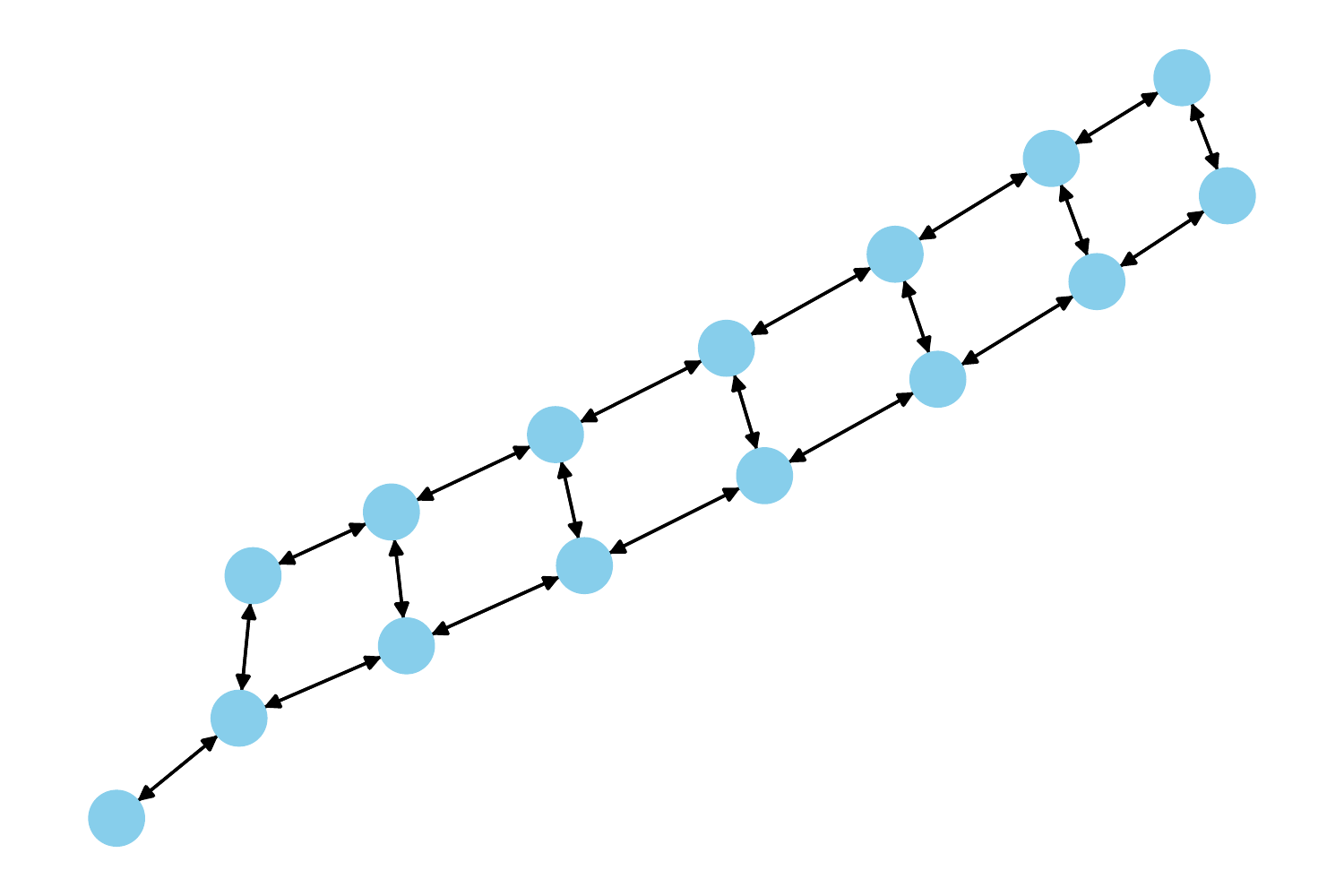}
		\caption{\melb{}}
		\label{fig:ibmq_16_melbourne coupling map}
	\end{subfigure}
	\hfill
	\begin{subfigure}[b]{0.24\textwidth}
		\centering
		\includegraphics[width=\textwidth]{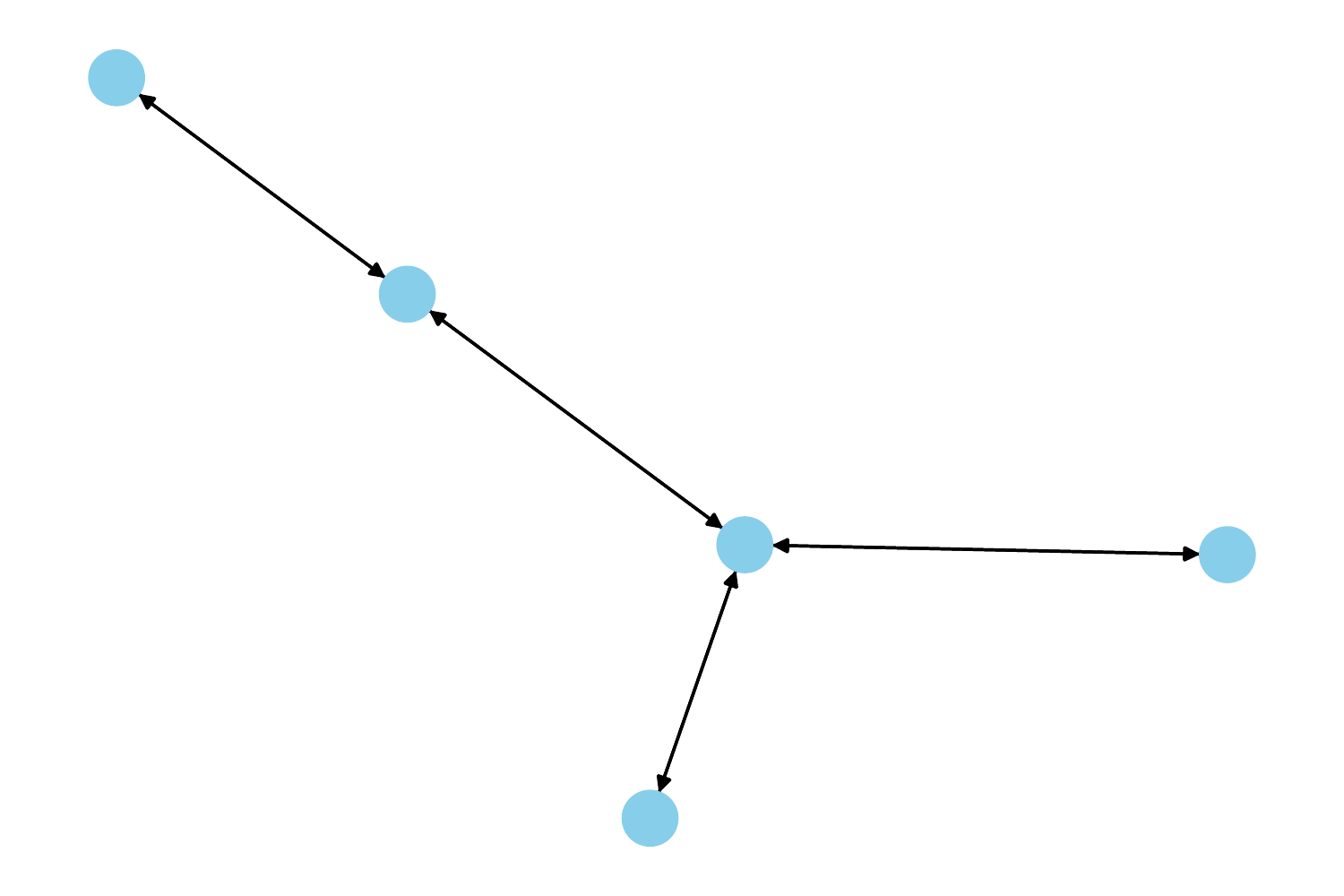}
		\caption{\our{}}
		\label{fig:ibmq_ourense coupling map}
	\end{subfigure}
    \caption{\textbf{Coupling maps of the devices studied in this work.}
    Vertices, represented by blue circles, correspond to qubits, while edges
    are directed from the control to the target qubits of permitted two-qubit
    gates.}
	\label{fig:coupling maps}
\end{figure*}

A \emph{coupling map} of a device is a graphical representation of how
two-qubit gates can be applied across the device. In this representation, each
qubit is represented by a vertex, with directed edges joining qubits between
which a two-qubit gate can be applied. For the devices considered here, this
two-qubit gate is a \CX{} gate, implemented using the cross-resonance
interaction of transmon qubits \cite{magesan2018effective}. The direction of
the edge is from the control to the target qubit of the \CX{} gate, with
bi-directional edges indicating that both qubits can be used as either the
control or target. The coupling maps of the devices investigated in this work
are shown in \figref{fig:coupling maps}. For those devices all edges are
bi-directional, although this is not typical when the asymmetric \CX{} is
employed.

As discussed in \secref{sec:stack}, a trade-off exists between the connectivity
of the device and the number of two-qubit gates necessary to implement a given
circuit. More highly connected coupling maps typically require fewer two-qubit
gates to implement a fixed unitary than less connected ones, owing to the
reduced need for \SWAP{} gates to account for discrepancies between the
coupling maps of the uncompiled circuit and the device. While this reduced
depth can reduce the impact of time based noise channels, this is
counterbalanced by the higher levels of cross-talk experienced by qubits
corresponding to vertices with high degree in the device's coupling map
\cite{PhysRevX.10.011022}.

\subsection{Device Calibration Information}
\label{app:noise levels}

\begin{figure*}
	\begin{subfigure}[b]{\textwidth}
		\centering
		\begin{tikzpicture}
			\node (graph) at (0,0)
			{\includegraphics[width=0.95\textwidth]{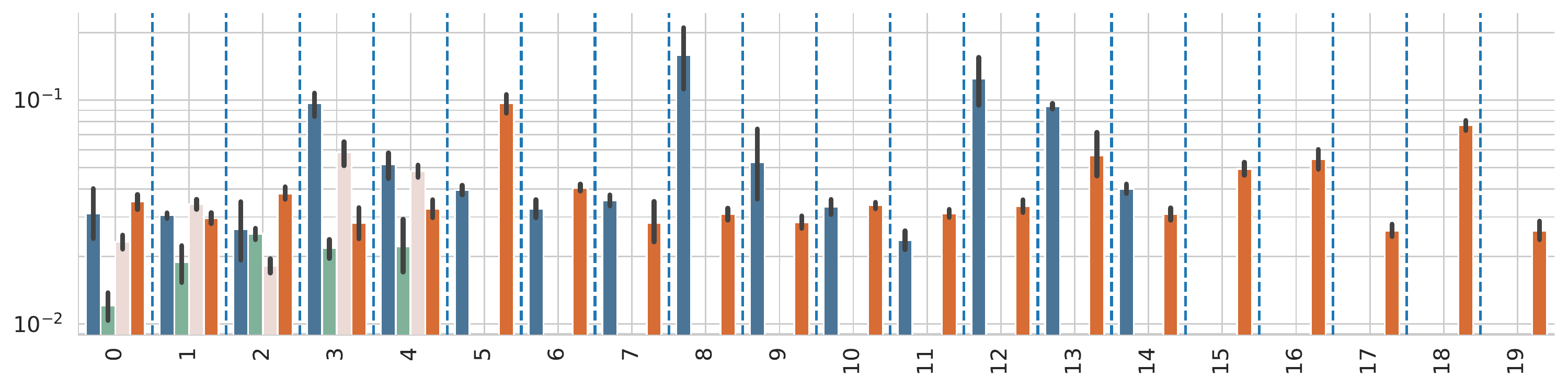}};

			\node at (graph.south) {Qubit};

			\node[rotate = 90] at (graph.west) {Error Rate};
		\end{tikzpicture}
		\caption{\textbf{Average readout error.} The readout error
		is the probability the state of a given qubit is incorrectly
		labelled.}
		\label{fig:full readout error}
	\end{subfigure}
	\begin{subfigure}[b]{\textwidth}
		\centering
		\begin{tikzpicture}
			\node (graph) at (0,0)
			{\includegraphics[width=0.95\textwidth]{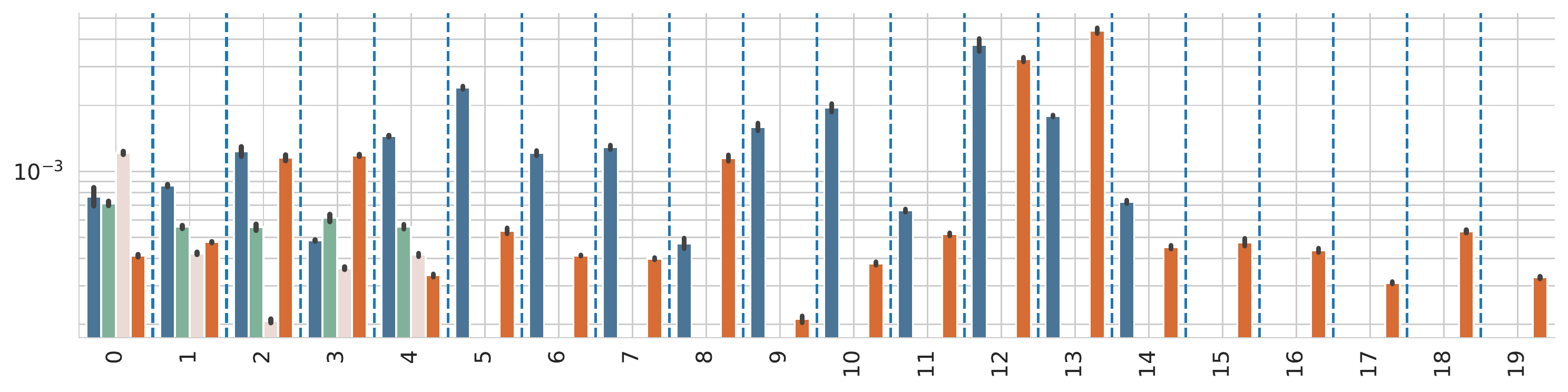}};

			\node at (graph.south) {Qubit};

			\node[rotate = 90] at (graph.west) {Error Rate};
		\end{tikzpicture}
		\caption{\textbf{Average error per $\Utwo$ gate.} The error per gate is a
		measure of how accurately the $\Utwo$ gate is applied.}
		\label{fig:full u2 error}
	\end{subfigure}
	\caption{\textbf{Error per single qubit operations on the devices used
	in this work.} Bars indicate the average error rates; error bars are one
	standard deviation. Data aggregated based on calibration data collected
	over the course of our experiments. Devices shown here are: \york{}
	\crule{ibmqx2}, \our{} \crule{ibmq_ourense}, \sing{}
	\crule{ibmq_singapore}, \melb{} \crule{ibmq_16_melbourne}. A logarithmic
	scale is used.}
	\label{fig:1 qubit noise levels}
\end{figure*}

\begin{figure*}
	\centering
	\begin{tikzpicture}
		\node[anchor=south west] (graph_even) at (0,0)
		{\includegraphics[width=0.95\textwidth]{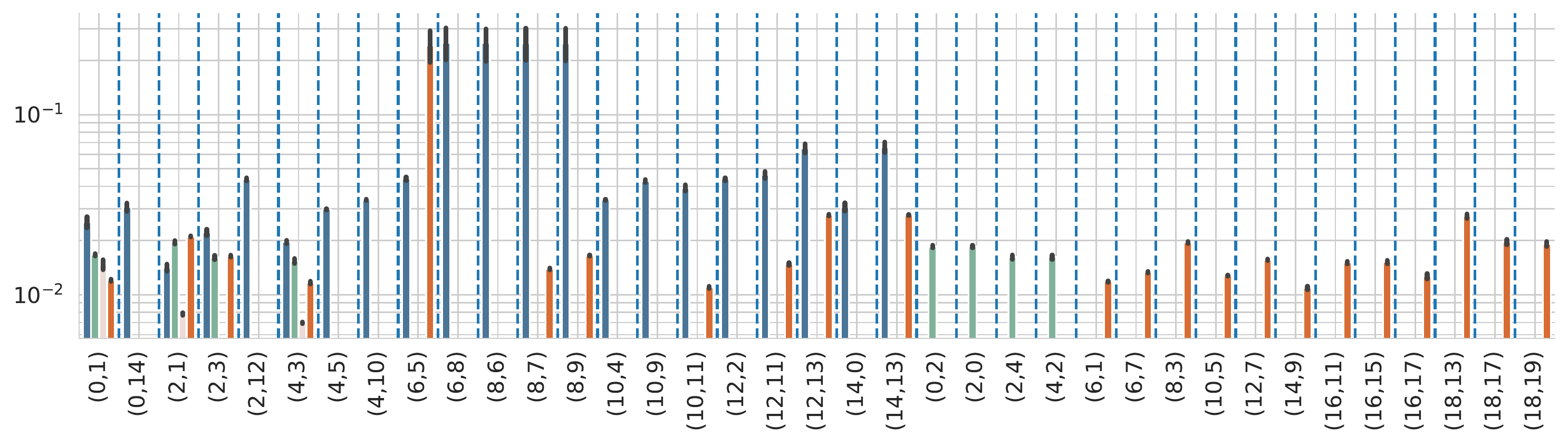}};

		\node[rotate = 90, anchor = south] at (0,0) {Error Rate};

		\node[anchor = north west] (graph_odd) at (0,0)
		{\includegraphics[width=0.95\textwidth]{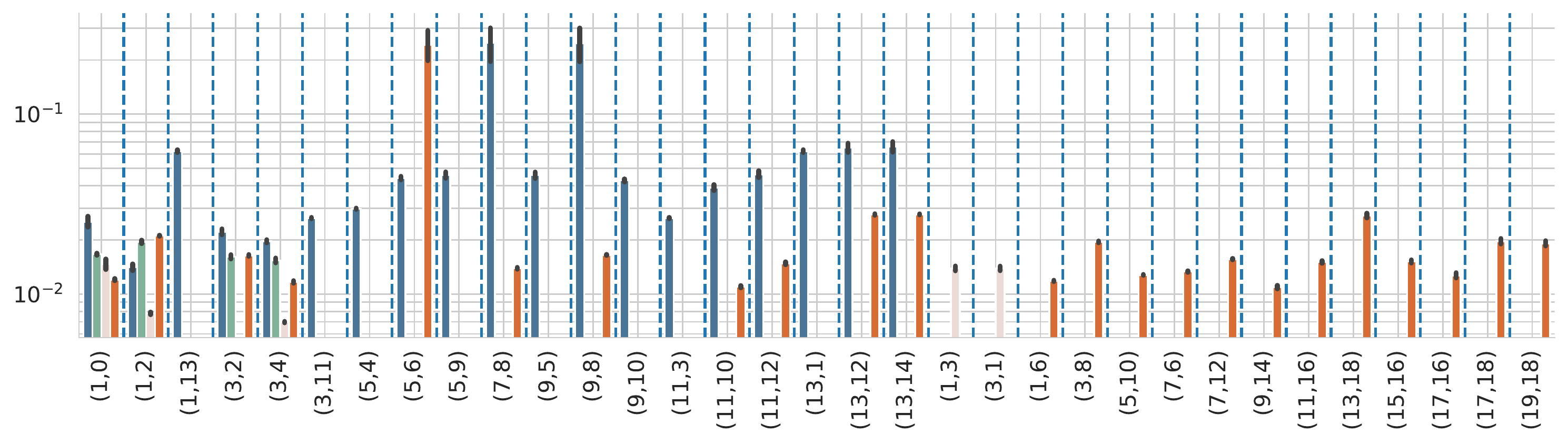}};

		\node at (graph_odd.south) {Qubits};
	\end{tikzpicture}
	\caption{\textbf{Average error per \CX{} operation on the devices used in
	this work.} The error per \CX{} gate is a measure of how accurately the
	\CX{} gate is applied. Bars indicate the average error rates; error bars
	are one standard deviation. Data aggregated based on calibration data
	collected over the course of our experiments. Devices shown here are:
	\york{} \crule{ibmqx2}, \our{} \crule{ibmq_ourense}, \sing{}
	\crule{ibmq_singapore}, \melb{} \crule{ibmq_16_melbourne}. A logarithmic
	scale is used.}
	\label{fig:full cx error}
\end{figure*}

The noise-aware tools employed by the compilation strategies explored in this
work consider three kinds of errors which can occur, namely: readout error,
single-qubit gate error, and two-qubit gate error. For the devices provided
through IBM Quantum, this information is contained in calibration data which is
accessible using tools in the \qiskit{} library, and is updated twice daily.
The experiments in this paper were conducted between 2020-01-29 and 2020-02-10
with the calibration data in \figref{fig:1 qubit noise levels} and
\figref{fig:full cx error} aggregated over this time period.

An assignment or readout error corresponds to an incorrect reading of the state
of the qubit; for example, returning ``0'' when the proper label is ``1'', or
vice-versa.  The probability of incorrectly labelling the qubit is called the
\emph{readout error}, denoted, $\epsilon^{\mathrm{a}}$, and is calculated as
\begin{equation}
	\epsilon^{\mathrm{a}} =
		\frac{\mathrm{Pr}\text{(``0''}|\ket{1})+\mathrm{Pr}\text{(``1''}|\ket{0})}{2}.
\end{equation}
$\epsilon^{\mathrm{a}}$ is estimated by repeatedly preparing a qubit in a known
state, immediately measuring it, and then counting the number of times the
measurement returns the wrong label. This value, for the devices explored in
this paper, is reported in \figref{fig:full readout error}.

Errors affecting the gates of the device correspond to an incorrect operation
applied by the device. There are many ways to quantify the effect of this
error, with IBM Quantum's devices reporting randomized benchmarking (RB)
numbers \cite{proctor2017what, randomized2008knill}. The RB number,
$\epsilon^{\mathrm{C}}$, is estimated by running many self-inverting Clifford
circuits, consisting of $m$ layers of gates drawn from the $n$-qubit Clifford
group, inverted at layer $m+1$. The \emph{survival probability}, which is the
probability the input state is unchanged, can then be estimated. Under a broad
set of noise models and assumptions \cite{proctor2017what,
Carignan_Dugas_2018}, this survival probability can be shown to decay
exponentially with $m$.  Consequently, it can be estimated by fitting a decay
curve of the form $Ap^{m}+B$. The RB number is related to $p \in
\sqrbrac{0,1}$, called the \emph{depolarisation/decay rate}, by
\begin{equation}
	\epsilon^{\mathrm{C}} = \brac{1-p} \brac{1-1/D},
\end{equation}
where $D = 2^{n}$, and $n$ is the number of qubits acted on by the Clifford
gates. $\epsilon^{\mathrm{C}}$, which is also referred to as the \emph{error
per Clifford} of the device, is minimised at $p=1$, in which case the survival
probability is constant and set by the state preparation and measurement
errors.

The Clifford gates necessary for RB must be compiled to the native gate set of
the device. Using an estimate of $\epsilon^{\mathrm{C}}$, an estimate of the
\emph{error per gate}, $\epsilon^{\mathrm{g}}_{G}$, for a gate $G$, can be
obtained by multiplying $\epsilon^{\mathrm{C}}$ by a factor related to the
average number of uses of $G$ when implementing a random Clifford operation:
\begin{equation}
	\epsilon^{\mathrm{g}}_{G} \sim \epsilon^{\mathrm{C}}\times \text{\# uses
	of}~G~\text{per Clifford}.
\end{equation}
Values for $\epsilon^{\mathrm{g}}_{\Utwo}$, the error per gate for $\Utwo$
gates, can be found in \figref{fig:full u2 error}, and
$\epsilon^{\mathrm{g}}_{\CX{}}$, that for \CX{} gates, in \figref{fig:full cx
error}. The commonly reported average fidelity for $\Uthree$ gates is
$1-\brac{1-\epsilon^{\mathrm{g}}_{\Utwo}}^2$.

There are many variants of randomized benchmarking, such as direct RB
\cite{proctor2019direct}, simultaneous RB \cite{gambetta2012characterization},
and correlated RB \cite{mckay2020correlated}. For details on the randomized
benchmarking protocol used by IBM Quantum, see \cite{magesan2011scalable,
magesan2012characterizing, corcoles2013process, gambetta2012characterization,
mckay2019three}.

The experiments necessary for cross-entropy benchmarking may themselves also be
used to estimate a depolarisation rate in a similar way to RB \cite{Arute2019}.
Instead of using random Clifford circuits, however, the random circuits are
run.  Under the assumption that the action of a random circuit can be described
using a depolarising error model (with equal-probability Pauli errors), then
the Pauli error, $\epsilon^{\mathrm{P}}$, can be estimated as
\begin{equation}
	\epsilon^{\mathrm{P}} = (1-p)(1-1/D^2).
\end{equation}
Here, $p$ is the depolarisation rate of the survival probability under the
action of random circuits, estimated as above. Interestingly,
$\epsilon^{\mathrm{P}}$ can be estimated using single and two-qubit RB
information.

Several important noise channels, most notably cross-talk, are not included in
the device calibration data. As shown in \secref{sec:results}, the effects of
this noise can be inferred through the application-motivated benchmarks we
introduce in this work, by showing the trade-off between connectivity of the
device and cross-talk \cite{PhysRevX.10.011022}.

\section{Empirical Relationship Between Heavy Output Generation Probability and
$\ell_1$-Norm Distance}
\label{app:l1_hog}

As discussed in \secref{sec:metric comparison}, the theoretical foundations for
believing that implementing \shallow{} to within a fixed \lone{} constitutes a
demonstration of \supremacy{} are stronger than for implementations with high
heavy output generation probability. That said,
\figref{fig:arch_CQC-noise_l1_shallow} and
\figref{fig:arch_CQC-noise_HO_shallow} contain similar features. For example,
\melb{} consistently performs the worst, with \sing{} and \our{} performing the
best by both figures of merit.  How then do these two figures of merit
generally relate to one another.

\newcommand{\normHO}{the normalised heavy output generation probability}

If the \lone{} was $0$, the experimental outcome frequencies would equal the
ideal outcome probabilities. Consequently, the heavy output probabilities would
be the same between the device and an ideal quantum computer. Because the heavy
output probability depends on the circuit in question, it is useful to
normalise the latter by the heavy output probability of an ideally-implemented
circuit. We define \emph{\normHO{}} as the ratio of the heavy output
probability of the device and the heavy output probability from an ideal
quantum computer. Hence if the \lone{} was 0, \normHO{} would be 1.

As the \lone{} increases, the  experimental frequencies increasingly differ
from the ideal outcome  probabilities. Two things then may happen: heavy
outputs are produced more regularly, in which case \normHO{} will grow above
$1$; or less regularly, in which case \normHO{} will fall below $1$. In
practice, we expect the distribution produced by the device to converge to the
uniform one as the noise increases, so we expect \normHO{} to fall with
increasing \lone{}.

\begin{figure*}

	\centering

	\begin{tikzpicture}

		\node (graph) at (0,0)
			 {\includegraphics[width=0.85\textwidth]{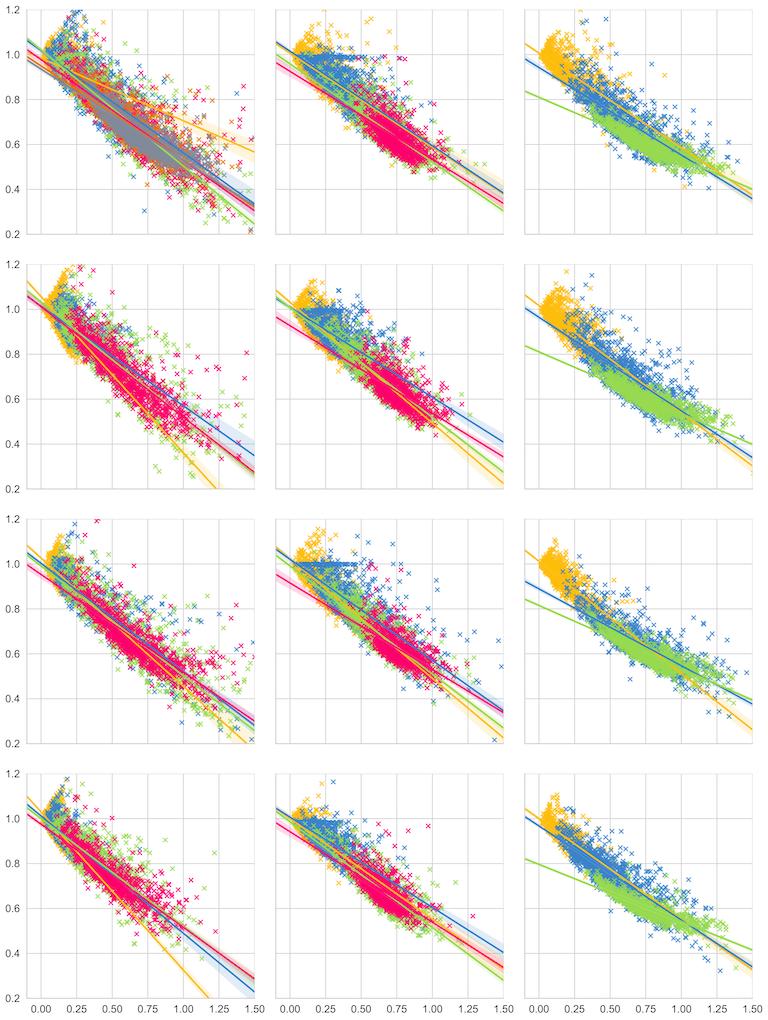}};

		\node[rotate=90, yshift=5pt] at (graph.west) {Normalised Heavy
		Output Probability};
		\node[yshift=-5pt] at (graph.south) {\lone{}};

		\node[yshift=15pt] at (graph.north) {\textbf{Circuit Class}};

		\node at ($(graph.north west)!1/6!(graph.north east)$)
		{\titlecap{\shallow{}}};
		\node at ($(graph.north west)!3/6!(graph.north east)$)
		{\titlecap{\square{}}};
		\node at ($(graph.north west)!5/6!(graph.north east)$)
		{\titlecap{\deep{}}};

		\node[xshift=15pt, rotate=270] at (graph.east) {\textbf{Device}};

		\node[rotate=270] at ($(graph.south east)!1/8!(graph.north
		east)$) {\our{}};
		\node[rotate=270] at ($(graph.south east)!3/8!(graph.north
		east)$) {\melb{}};
		\node[rotate=270] at ($(graph.south east)!5/8!(graph.north
		east)$) {\york{}};
		\node[rotate=270] at ($(graph.south east)!7/8!(graph.north
		east)$) {\sing{}};

	\end{tikzpicture}
	\caption{\textbf{Scatter plot and linear regression line comparing
	\normHO{} and \lone{}}. Each point corresponds to one circuit of the
	class and width as labelled. Colours correspond to numbers of qubits in
	the following way: 2 \crule{2}, 3 \crule{3}, 4 \crule{4}, 5 \crule{5}, 6
	\crule{6}, 7 \crule{7}.}
	\label{fig:correlation}
\end{figure*}

The empirical relationship between \normHO{} and \lone{} is shown in
\figref{fig:correlation}. For each circuit, \figref{fig:correlation} plots the
\lone{} of the distribution produced by a real device against \normHO{}. As
expected, a negative correlation exists between these two figures of merit. For
the deepest circuits, and in particular the widest circuits from the \deep{}
class, the cluster of points indicates that the \normHO{} falls more slowly as
the \lone{} becomes larger. This is because the minimum value of heavy output
generation probability is being reached, which is to say that the output
distribution from the real device has converged to the uniform one, while more
detail can be extracted by considering the \lone{}.

This correlation is encouraging as, in the regime where it becomes impossible
to calculate the \lone{}, we can be justified in believing the correlation
between the features present in \secref{sec:results} persist. This is to say
that we may be able to make lose predictions about the performance of a
\stack{} as measured by \lone{} based on it's performance as measured by
\HOG{}. This reasoning is similar to that used when \XE{} is used to predict
demonstrations of \supremacy{} in the regime when it becomes impossible to
calculate \cite{Arute2019}. 

\end{document}